\documentclass{aa}  

\usepackage{graphicx}
\usepackage{txfonts}
\usepackage{graphicx}	% Including figure files
\usepackage{amsmath}	% Advanced maths commands
\usepackage{amssymb}	% Extra maths symbols
\usepackage{natbib}
\usepackage[para]{threeparttable}
\usepackage{verbatim}
\usepackage{mathtools}
\usepackage{xcolor}

\begin{document}

   \title{Extension of the HCOOH and CO$_2$ solid-state reaction network during the CO freeze-out stage: inclusion of H$_2$CO}

    \author{D. Qasim\inst{1},
   T. Lamberts\inst{2},
  J. He\inst{1},
   K.-J. Chuang\inst{1}\thanks{Present address: Laboratory Astrophysics Group of the Max Planck Institute for Astronomy at the Friedrich Schiller University Jena,
Institute of Solid State Physics, Helmholtzweg 3, D-07743 Jena, Germany},
   G. Fedoseev\inst{3},
   S. Ioppolo\inst{4},
   A.C.A. Boogert\inst{5},
          \and
          H. Linnartz\inst{1}
          }

   \institute{Sackler Laboratory for Astrophysics, Leiden Observatory, Leiden University, PO Box 9513, NL--2300 RA Leiden, The Netherlands\\
              \email{dqasim@strw.leidenuniv.nl}
             \and Leiden Institute of Chemistry, Leiden University, PO Box 9502, NL--2300 RA Leiden, The Netherlands \
             \and INAF--Osservatorio Astrofisico di Catania, via Santa Sofia 78, 95123 Catania, Italy
             \and School of Electronic Engineering and Computer Science, Queen Mary University of London, Mile End Road, London E1 4NS, UK
             \and Institute for Astronomy, University of Hawaii at Manoa, 2680 Woodlawn Drive, Honolulu, HI 96822--1839 
             }

   \date{Received X; accepted Y}

% \abstract{}{}{}{}{} 
% 5 {} token are mandatory
 
  \abstract
   {Formic acid (HCOOH) and carbon dioxide (CO$_2$) are simple species that have been detected in the interstellar medium. The solid-state formation pathways of these species under experimental conditions relevant to prestellar cores are primarily based off of weak infrared transitions of the HOCO complex and usually pertain to the H$_2$O-rich ice phase, and therefore more experimental data are desired.}
   {Here, we present a new and additional solid-state reaction pathway that can form HCOOH and CO$_2$ ice at 10 K `non-energetically' in the laboratory under conditions related to the "heavy" CO freeze-out stage in dense interstellar clouds, i.e., by the hydrogenation of an H$_2$CO:O$_2$ ice mixture. This pathway is used to piece together the HCOOH and CO$_2$ formation routes when H$_2$CO or CO reacts with H and OH radicals.}
   {Temperature programmed desorption - quadrupole mass spectrometry (TPD-QMS) is used to confirm the formation and pathways of newly synthesized ice species as well as to provide information on relative molecular abundances. Reflection absorption infrared spectroscopy (RAIRS) is additionally employed to characterize reaction products and determine relative molecular abundances.}
   {We find that for the conditions investigated in conjunction with theoretical results from the literature, H + HOCO and HCO + OH lead to the formation of HCOOH ice in our experiments. Which reaction is more dominant can be determined if the H + HOCO branching ratio is more constrained by computational simulations, as the HCOOH:CO$_2$ abundance ratio is experimentally measured to be around 1.8:1. H + HOCO is more likely than OH + CO (without HOCO formation) to form CO$_2$. Isotope experiments presented here further validate that H + HOCO is the dominant route for HCOOH ice formation in a CO-rich CO:O$_2$ ice mixture that is hydrogenated. These data will help in the search and positive identification of HCOOH ice in prestellar cores.}
   {}


   \authorrunning{Qasim et al.}
   \titlerunning{Constraints on the formation of HCOOH and CO$_2$ ice}
   
   \maketitle
%
%________________________________________________________________

\section{Introduction}

HCOOH and CO$_2$ have been detected across various environments in space. HCOOH has been confirmed in the gas-phase \citep{zuckerman1971microwave,winnewisser1975detection,irvine1990detection,turner1999physics,liu2001observations,ikeda2001survey,liu2002formic,requena2006organic,bottinelli2007hot,taquet2017chemical,favre2018first}, but its identification in the solid-state is uncertain \citep{boogert2015observations}. Abundances relative to H$_2$O ice of less than 0.5 to up to 6 percent towards dense clouds, and low and high mass Young Stellar Objects have been derived \citep{schutte1999weak,gibb2004interstellar,bisschop2007infrared,oberg2011spitzer,boogert2015observations}, providing the identification of HCOOH is indeed correct. CO$_2$ has been detected in the gas-phase \citep{dartois1998molecular,van1998can,boonman2003gas} and abundantly in the solid-state \cite[and references therein]{d1989discovery,van1996search,de1996sws,whittet1998detection,gerakines1999observations,pontoppidan2008c2d,cook2011thermal,kim2012co2,poteet2013anomalous,ioppolo2013solid,boogert2015observations,suhasaria2017solid}, and is one of the most ubiquitous ice constituents in the interstellar medium. Here, the observed CO$_2$:H$_2$O ice ratios are 15-35\%.     

At astrochemically relevant temperatures of $\leq$ 20 K, HCOOH ice has been shown to be formed by proton irradiation of H$_2$O and CO \citep{hudson1999laboratory}, electron irradiation of H$_2$O and CO \citep{bennett2010laboratory}, UV irradiation of H$_2$O and CO \citep{watanabe2007laboratory}, hydrogenation of CO and O$_2$ \citep{ioppolo2010surface}, and a combined UV irradiation and hydrogenation of H$_2$O and CO \citep{watanabe2007laboratory}. Upper limit values or tentative identification of HCOOH formation in UV-induced experiments containing CH$_3$OH \citep{oberg2009formation,paardekooper2016laser} and H$_2$CO \citep{butscher2016radical} have also been reported. CO$_2$ ice can also be produced by both, `energetic' and `non-energetic' processes, where `non-energetic' refers to a radical-induced process without the involvement of UV, cosmic rays, and/or other `energetic' particles. For `energetic' processes, CO$_2$ has been observed and formed experimentally by UV-irradiation of CO-containing ices \citep{gerakines1996ultraviolet,ehrenfreund1997infrared,cottin2003photodestruction,loeffler2005co}, electron-induced radiation \citep{jamieson2006understanding,martin2008chemistry,bennett2009mechanistical}, as well as through ion bombardment \citep{moore1996infrared,palumbo1998profile,satorre2000co,trottier2004carbon}. As relevant to this study, CO$_2$ has also been shown to be formed from the irradiation of CO:O$_2$ ices \citep{satorre2000co,strazzulla1997possible,bennett2009experimentalCO}.

Solid-state laboratory experiments suggest that interstellar HCOOH and CO$_2$ may have an icy origin and with common formation pathways. CO$_2$ is typically formed alongside HCOOH \citep{hudson1999laboratory,ioppolo2010surface,bennett2010laboratory,butscher2016radical} or from HCOOH \citep{andrade2013chemical,bergantini2013processing,ryazantsev2015radiation}. It has also been formed without co-detection of HCOOH \citep{oba2010experimental,oba2010formation,raut2011solid,ioppolo2011surface,ioppolo2013solid,ioppolo2013surfreside2,raut2012radiation,minissale2015solid,suhasaria2017solid}, although not all the studies explicitly state a non-detection of HCOOH or involve hydrogen. And in the case of the HOCO intermediate, it can dissociate upon hydrogenation to form CO$_2$ rather than hydrogenate to form HCOOH \citep{ioppolo2011surfacerend,linnartz2015atom}.     

While the bulk of CO$_2$ and probably HCOOH is formed early in the cloud evolution, alongside with H$_2$O, the aforementioned laboratory experiments indicate that some HCOOH and CO$_2$ formation may occur during the heavy CO freeze-out stage as well. This is additionally supported by observations of CO$_2$ ice in a CO environment \citep{pontoppidan2008c2d}, where the fraction of CO$_2$ in a CO environment is 15-40\%. The authors suggest that a quiescent formation mechanism could be at play for CO$_2$ formation, in addition to cosmic-ray induced chemistry. Such a mechanism may arise from the hydrogenation of a CO:O$_2$ ice mixture, as some O$_2$ is likely mixed with CO at greater cloud depths \citep{tielens1982model,vandenbussche1999constraints,qasim2018formation}. Additionally, OH radicals near the top of the H$_2$O-rich ice may interact with incoming CO molecules. During the heavy CO freeze-out stage, a considerable amount of CO freezes out \citep{pontoppidan2006spatial,boogert2015observations,qasim2018formation} and can be hydrogenated to form H$_2$CO, CH$_3$OH, and more complex organics \citep{chuang2015h,fedoseev2017formation}. The solid-state chemistry during the CO freeze-out stage is primarily driven by atom addition and abstraction reactions \citep{linnartz2015atom}. 

The experimentally and/or theoretically investigated `non-energetic' formation pathways for HCOOH and CO$_2$ ice that are reported in the literature are: 
\begin{alignat}{2}%\label{eq1}
\label{eq01}
\mathrm{HCO + OH} &\rightarrow \mathrm{HCOOH}\\
\label{eq1a}
\mathrm{OH + CO} &\rightarrow \mathrm{HOCO}, \mathrm{H + HOCO} &&\rightarrow \mathrm{HCOOH}\\
\label{eq1c}
\mathrm{OH + CO} &\rightarrow \mathrm{HOCO}, \mathrm{H + HOCO} &&\rightarrow \mathrm{H_2 + CO_2}\\
\label{eq1d}
\mathrm{OH + CO} &\rightarrow \mathrm{HOCO}, \mathrm{OH + HOCO} &&\rightarrow \mathrm{H_2O + CO_2}\\
\label{eq2}
\mathrm{OH + CO} &\rightarrow \mathrm{H + CO_2}\\
\label{eq3}
\mathrm{CO + O} &\rightarrow \mathrm{CO_2}\\
\label{eq4}
\mathrm{H_2CO + O} &\rightarrow \mathrm{CO_2 + H_2}
\end{alignat}

Although investigated in detail before, many of the listed routes to HCOOH and CO$_2$ formation are yet to have tight experimental constraints. To our knowledge, only one study showed the formation of HCOOH under `non-energetic' conditions \citep{ioppolo2010surface}, and within this study, the reaction pathway involving the HOCO complex was considered. HCO + OH was reported to be unlikely since the HCO derivatives, H$_2$CO and CH$_3$OH, went undetected in the RAIR experiments under their experimental conditions. However, models suggest that HCO + OH is relevant to the solid-state chemistry of the prestellar core phase \citep{garrod2006formation}. \citet{oba2010experimental} also studied the reaction of OH and CO at low temperatures under different experimental conditions than those used by \citet{ioppolo2010surface}, and did not positively identify HCOOH formation. For the formation routes of CO$_2$, particularly reactions \ref{eq1c}, \ref{eq1d}, and \ref{eq2} are of concern as it is uncertain which route dominates within an experiment, and thus which one is expected to be relevant to the interstellar medium (ISM). The reactant HOCO was considered an ingredient for CO$_2$ formation due to the reported weak IR band(s) of the HOCO complex \citep{oba2010experimental,ioppolo2011surface}. In the work by \citet{noble2011co2}, formation of CO$_2$ + H by solid-state OH + CO was reported to occur under their experimental conditions. 

\begin{table*}
	\centering
	\caption{A detailed list of the experiments described in this study. Fluxes are calculated by the Hertz-Knudsen equation except for the H/D flux.}
    \label{table1}
    \begin{threeparttable}
	\begin{tabular}{c c c c c c c c c c} % 10 columns, alignment for each
		\hline
		No. & Experiments & T$_{\mathrm{sample}}$ & Flux$_{{\mathrm{{H}_2{CO}}}/{\mathrm{CO}}}$ & Flux$_{\mathrm{H/D}}$ & Flux$_{\mathrm{O}_2}$ & Flux$_{\mathrm{{H}_2{O}}}$ & Flux$_{\mathrm{HCOOH}}$ & Time \\
 & & (K) & (cm$^{-2}$s$^{-1}$) & (cm$^{-2}$s$^{-1}$) & (cm$^{-2}$s$^{-1}$) & (cm$^{-2}$s$^{-1}$) & (cm$^{-2}$s$^{-1}$) & (s) \\
 
		\hline
		1 & H$_{2}$CO + H + O$_2$ & 10 & $3 \times 10^{12}$ & $5 \times 10^{12}$ & $4 \times 10^{12}$ & - & - & 43200\\ %07062017(1)
        2 & H$_{2}$CO + H + O$_2$ & 10 & $3 \times 10^{12}$ & $5 \times 10^{12}$ & $4 \times 10^{12}$ & - & - & 14400\\ %07062017(1)
		3 & H$_{2}$$^{13}$CO + H + O$_2$ & 10 & $3 \times 10^{12}$ & $5 \times 10^{12}$ & $4 \times 10^{12}$ & - & - & 14400\\ %07102017
        4 & H$_{2}$CO + H + $^{18}$O$_2$ & 10 & $3 \times 10^{12}$ & $5 \times 10^{12}$ & $4 \times 10^{12}$ & - & - & 14400\\ 
        5 & H$_{2}$$^{13}$CO + H + $^{18}$O$_2$ &  10 & $3 \times 10^{12}$ & $5 \times 10^{12}$ & $4 \times 10^{12}$ & - & - & 14400\\ %07142017
		6 & HCOOH & 10 & - & - & - & - & $7 \times 10^{11}$ & 14400\\ %07132017
        7 & H$_{2}$CO + O$_2$ + H$_2$O &  10 & $3 \times 10^{12}$* & - & $4 \times 10^{12}$ & $4 \times 10^{11}$* & - & 4800\\ %07032017(25)
        8 & H$_2$O &  15 & - & - & - & $2 \times 10^{13}$ & - & 2160\\ %01032018(23)
        9 & H + O$_2$ &  15 & - & $2 \times 10^{12}$ & $3 \times 10^{12}$ & - & - & 9120\\ %01032018(23)
        10$^{\ddagger}$ & CO + D + O$_2$ & 10 & $3 \times 10^{12}$ & $7 \times 10^{12}$ & $4 \times 10^{12}$ & - & - & 14400\\
        11$^{\ddagger}$ & CO + H + O$_2$ & 10 & $3 \times 10^{12}$ & $7 \times 10^{12}$ & $4 \times 10^{12}$ & - & - & 14400\\
               
\hline
\end{tabular}
	\begin{tablenotes}
	\item[*] Fluxes adjusted during deposition in order to achieve the desired RAIR integrated band area ratio between H$_2$CO and H$_2$O.\\
    \item[$^{\ddagger}$] H and D atoms are produced by the MWAS. Experiments 1-5 and 9 utilize the HABS.
	\end{tablenotes}
\end{threeparttable}
\end{table*}

In this paper, we propose an additional way to form HCOOH and CO$_2$ under conditions relevant to the heavy CO freeze-out stage, and that is through the hydrogenation of an H$_2$CO:O$_2$ ice mixture, where O$_2$ is used as a tool to produce an abundant amount of OH radicals in the ice \citep{cuppen2010water}. It is stressed that the focus of this work is not to mimic a realistic interstellar ice, but to identify potential interstellar ice reaction channels. We use this new experimental finding, deuterated experiments, theoretical results available from the literature, and revisit previous experimentally studied reactions to provide greater constraints on the HCOOH and CO$_2$ `non-energetic' formation pathways. Section~\ref{setup} overviews the current state of the experimental apparatus and details the experimental parameters and methods used. Section~\ref{results} discusses the formation pathways of HCOOH and CO$_2$. Section~\ref{astro} outlines in particular how this study can contribute to the search for HCOOH ice in dense clouds. Section~\ref{conc} lists the concluding remarks of this work.

\section{Methodology}
\label{setup}
\subsection{Experimental setup}

The experiments described here are performed with SURFRESIDE$^2$; an ultrahigh vacuum (UHV) system designed to investigate the atom-induced reaction dynamics that take place in interstellar ices found in dark clouds. Within the main chamber, a closed cycle helium cryostat is connected to a gold-plated copper sample, which acts as a platform for solid-state reactions. More details on the design of the setup can be found in \citet{ioppolo2013surfreside2}, and recent modifications to the setup are included in \citet{chuang2018reactive}. 

To date, atoms including H, D, N, and O can be formed by the combination of two atom beam lines. Each line is placed in its own vacuum chamber and connected to the main chamber through a shutter valve, which has a base pressure of $\sim$$10^{-10}$ mbar. Ample hydrogen and deuterium atoms are formed by a Hydrogen Atom Beam Source (HABS) \citep{tschersich1998formation,tschersich2000intensity,tschersich2008design} and a Microwave Atom Source (MWAS; Oxford Scientific Ltd.). For the HABS, atoms are formed by the thermal cracking of hydrogen molecules. Because the filament within the HABS reaches a temperature of 2065 K in this study, a nose-shaped quartz tube is placed at the end of the HABS in order to collisionally cool the H-atoms. Such a tube is also placed at the end of the MWAS, since the atoms are created by the bombardment of `energetic' electrons that are stimulated by a 2.45 GHz microwave power supply (Sairem) at 275 W. Both sources are utilized for two series of experiments (see Table~\ref{table1}), where the HABS is exploited for most of the experiments. The MWAS was used when the HABS became unavailable due to maintenance. 

Gases, liquids, and solids are used to make the necessary ice mixtures. H$_2$ (Linde 5.0) gas goes through the HABS and MWAS, while D$_2$ (Praxair 99.8\%) is only fragmented by the MWAS. O$_2$ (Linde Gas 99.999\%) and $^{18}$O$_2$ gases (Campro Scientific 97\%) flow through the vacuum chamber of the MWAS (microwave source off) and into the main chamber. HCOOH and H$_2$O liquids are connected to turbomolecular-pumped dosing lines and undergo freeze-pump-thaw cycles in order to get rid of volatile impurities. Paraformaldehyde (Sigma-Aldrich 95\%) and paraformaldehyde-$^{13}$C (Sigma-Aldrich 99\%) powders are also connected to the pre-pumped dosing lines, and are thermally decomposed by a bath of hot water in order to form H$_2$CO and H$_2$$^{13}$CO vapours, respectively. CO gas (Linde Gas 99.997\%) is also introduced through one of the dosing lines. Both dosing lines terminate with manually-operated leak valves. All isotopes are used for the purpose of constraining the newly formed ice species, HCOOH and CO$_{2}$, and their reaction pathways.

The techniques employed to investigate the ice composition, reaction pathways, and relative chemical abundances are temperature programmed desorption - quadrupole mass spectrometry (TPD-QMS) and reflection absorption infrared spectroscopy (RAIRS). For TPD-QMS, a quadrupole mass spectrometer (Spectra Microvision Plus LM76) is used to probe the mass-to-charge (\emph{m/z}) values of the sublimated ice reactants and products as a function of temperature. Resistive heating is used to heat the sample, which has a temperature range of 7-450 K. A silicon diode sensor placed at the back of the sample is used to measure the temperature and has an absolute accuracy of 0.5 K. An electron impact ionization energy of 70 eV is set for all experiments, while a TPD-QMS ramp rate of 2 or 5 K/min is chosen. It has been found that this small change in the ramp rate does not affect the main conclusions of this study. For RAIRS, a Fourier transform infrared spectrometer (FTIR; Agilent Cary 640/660) with a used wavenumber range of 4000-700 cm$^{-1}$ and a resolution of 1 cm$^{-1}$ is utilized to probe solid-state species through their molecular vibrations in the ice.The QMS detection limit is around 0.005 monolayers (equivalent to the amount in the solid-state), and for the FTIR, it is around an order of magnitude less for species with relatively high band strengths. 

\subsection{Experimental methods}
\label{methods}

Details of the experiments used for this study are outlined in Table~\ref{table1}. The H and D-atom fluxes are derived from an absolute D-atom flux that was determined by a QMS \citep{ioppolo2013surfreside2}. All other fluxes are calculated by the Hertz-Knudsen equation. The rationale for the set of experiments is discussed in the following paragraphs.

Experiments 1-11 are, in part, used to show the unequivocal results of HCOOH and CO$_2$ ice formation at 10 K. The OH radicals needed for the formation of HCOOH and CO$_2$ are formed by H + O$_2$ \citep{cuppen2010water}. All experiments involve the co-deposition technique (i.e., reactants are deposited simultaneously). Co-deposition is more representative of interstellar conditions and enhances the possibility of radical recombination reactions. The purpose of the lengthy experiment, 1, is to increase the product abundance in order to probe the formed ice species via RAIRS, which is a less sensitive technique compared to TPD-QMS under our specific experimental settings. The isotopes used in experiments 3-5 are applied to observe the \emph{m/z} and infrared band shifts in the TPD-QMS and RAIR data, respectively. These shifts are compared to the spectra of experiment 2, which represents the principal reaction of this study. Experiments 6-9 are used to confirm the HCOOH infrared signature in the RAIR data of experiment 1. Experiments 10-11 are used to validate the H + HOCO pathway in CO-rich ices, and thus provide additional insight into the formation of HCOOH and CO$_2$ in the H$_2$CO + H + O$_2$ experiment.

RAIR data are exploited to determine the HCOOH:CO$_2$ relative abundance using a modified Lambert-Beer equation. Band strength values of $5.4 \times 10^{-17}$ and $7.6 \times 10^{-17}$ cm molecule$^{-1}$ are used to determine the column densities of HCOOH and CO$_2$, respectively, and are obtained from \citet{bouilloud2015bibliographic}. Note that the values are multiplied by a transmission-to-RAIR proportionality factor, and the procedure to obtain this factor is discussed in \citet{qasim2018formation}. A HCOOH:CO$_2$ ice abundance ratio of 1.9:1 is measured. To check the validity of using RAIR data to measure abundances, the abundances are additionally determined from TPD-QMS data. For the TPD measurements of H$^{13}$COOH and $^{13}$CO$_2$, the peak heights are measured at 157 K for H$^{13}$COOH desorption (\emph{m/z} = 47), and 79 and 150 K for $^{13}$CO$_2$ desorption (\emph{m/z} = 45). The formula to determine the column density from TPD data is found in \citet{martin2015uv} and references therein. The total ionization cross sections used are $5.09 \times 10^{-16}$ cm$^{2}$ for H$^{13}$COOH \citep{mozejko2007calculations} and $2.74 \times 10^{-16}$ cm$^{2}$ for $^{13}$CO$_2$ \citep{orient1987electron}.
The sensitivity values are 0.162 and 0.176 for \emph{m/z} = 47
and 45, respectively, and are collected from SURFRESIDE$^{2}$ \citep{chuang2018formation}. A H$^{13}$COOH:$^{13}$CO$_2$ ice abundance ratio of 1.6:1 is determined. The discrepancy is within range for the uncertainties that are associated with each method. Particularly with RAIRS, the accuracy maybe less than that of TPD-QMS since RAIRS is more sensitive to surface features than by the bulk of the ice.

The relative abundances of other formed products are determined by RAIR data solely, since the TPD-QMS data has overlapping \emph{m/z} values (e.g., the CO$^{+}$ signal is completely dominated by the signal of the CO$^{+}$ fragment of H$_2$CO$^{+}$). Band strength values of $2.1 \times 10^{-17}$ cm molecule$^{-1}$ for H$_2$O$_2$, $7.6 \times 10^{-18}$ cm molecule$^{-1}$ for H$_2$O, and $5.2 \times 10^{-17}$ cm molecule$^{-1}$ for CO are used to determine the column densities of the three species, where the band strength values for H$_2$O$_2$ and H$_2$O are determined from \citet{loeffler2006decomposition}, and the value for CO is from \citet{chuang2018reactive}. An H$_2$O$_2$:H$_2$O:CO abundance ratio of 1:0.7:0.02 is measured.

\begin{figure}
\includegraphics[width=\columnwidth]{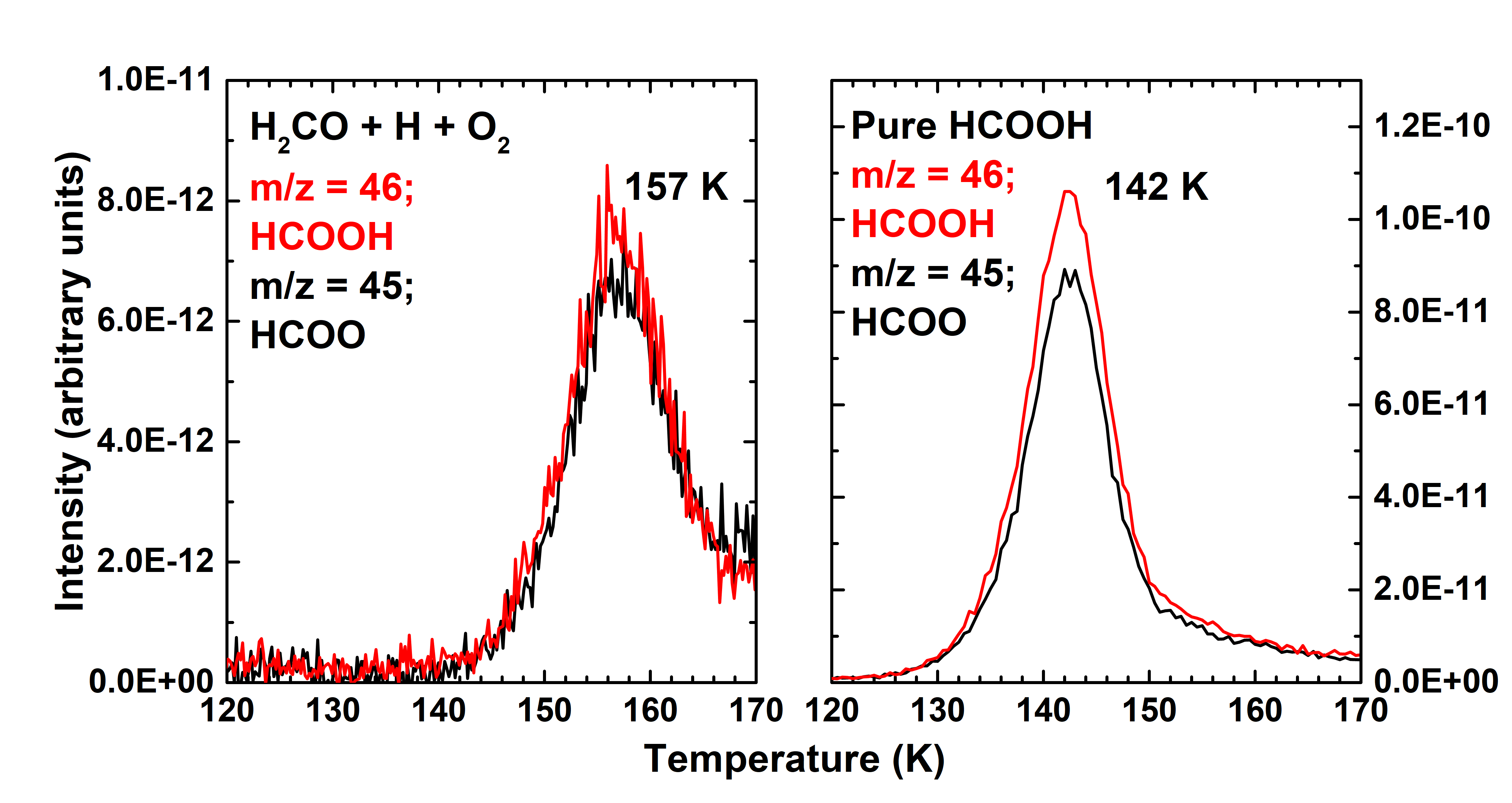}
\includegraphics[width=\columnwidth]{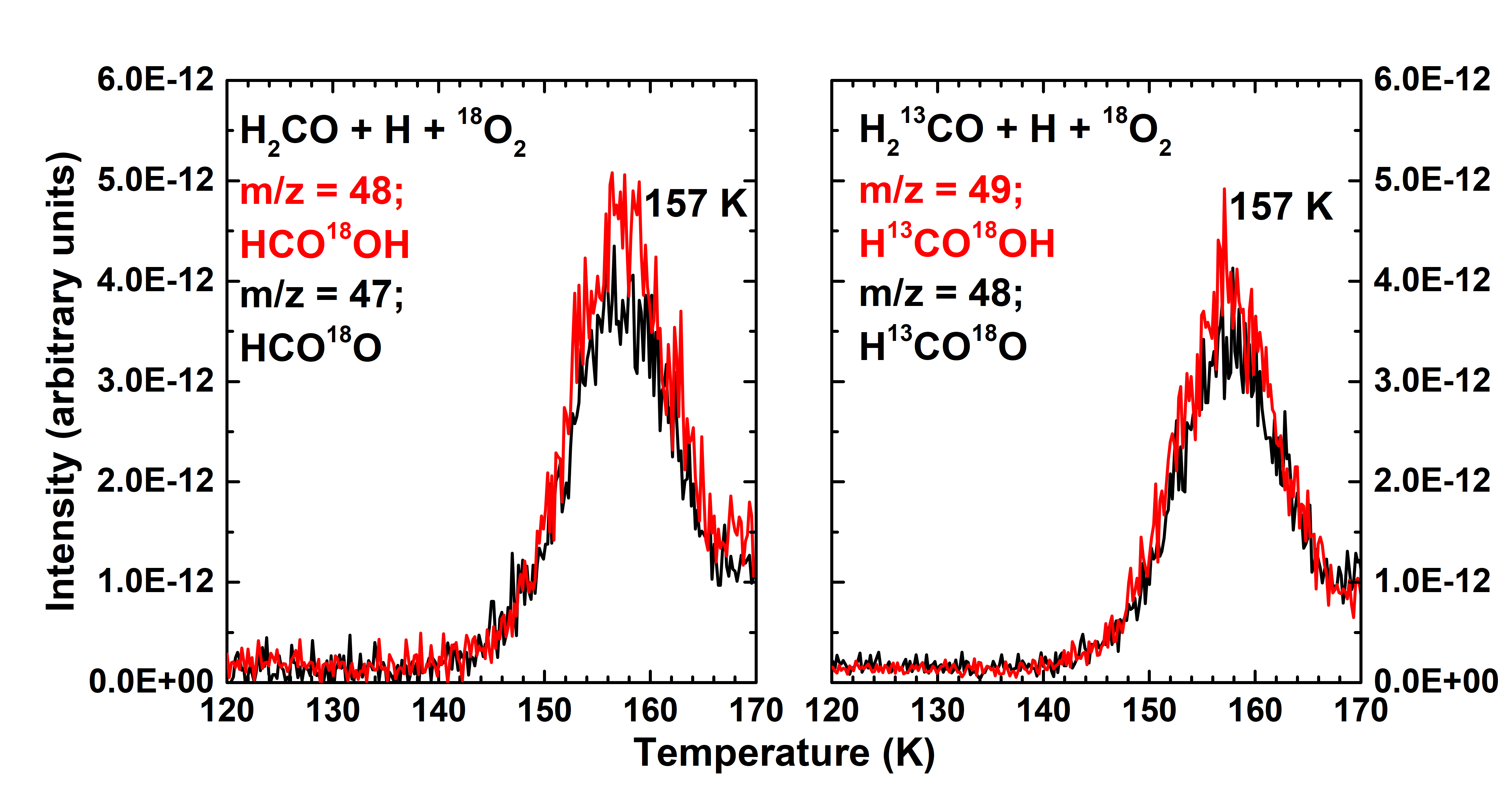}
\caption{TPD-QMS spectra of H$_{2}$CO + H + O$_2$ (top left; exp. 2) and HCOOH (top right; exp. 6). TPD-QMS spectra of H$_{2}$CO + H + $^{18}$O$_2$ (bottom left; exp. 4) and H$_{2}$$^{13}$CO + H + $^{18}$O$_2$ (bottom right; exp. 5). All spectra are recorded after ice growth at 10 K.}
\label{fig1}
\end{figure}

\begin{figure}
\includegraphics[width=\columnwidth]{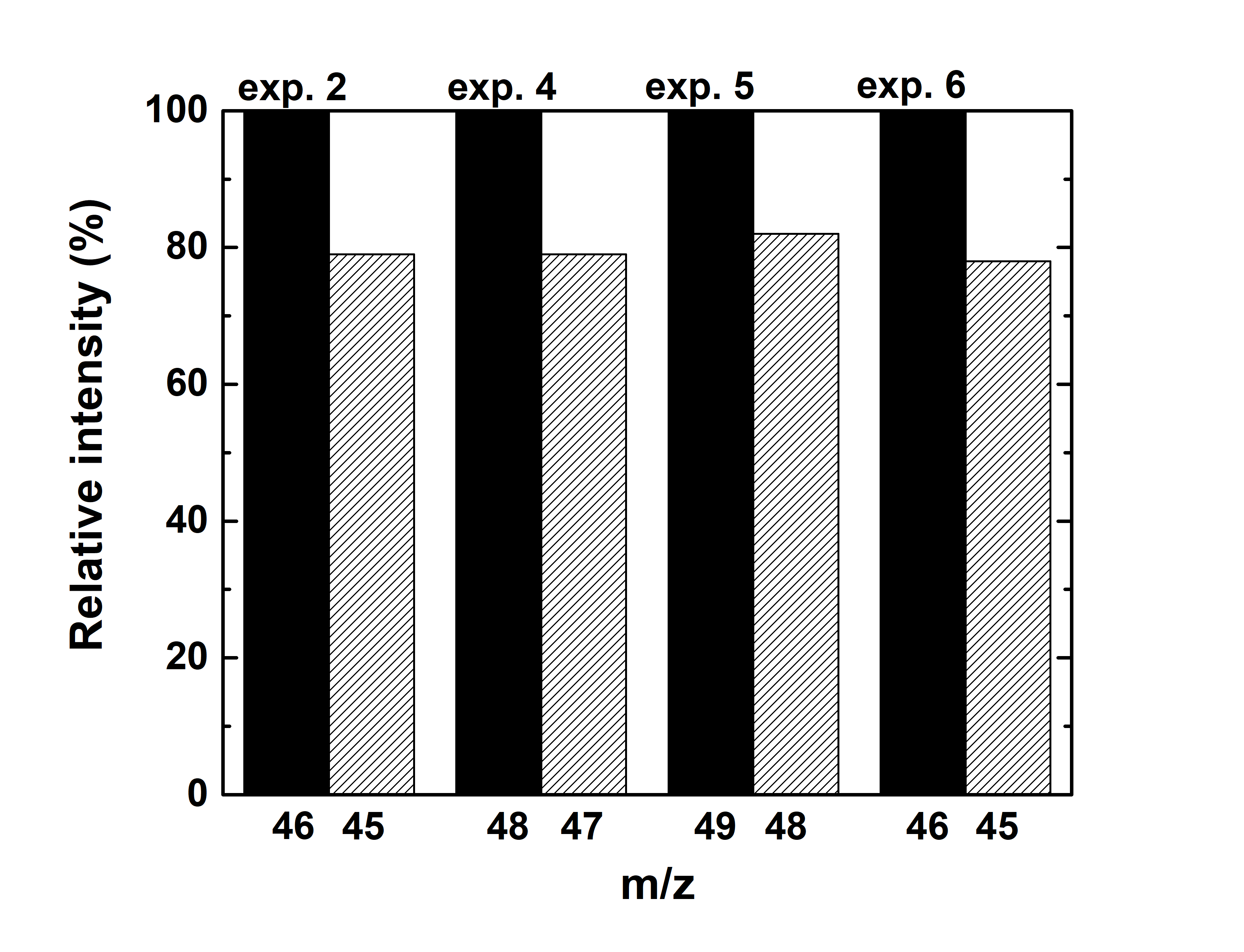}
\caption{The QMS cracking pattern of the desorption feature peaking at 157 K in the H$_2$CO + H + O$_2$ experiment (exp. 2), H$_{2}$CO + H + $^{18}$O$_2$ experiment (exp. 4), H$_{2}$$^{13}$CO + H + $^{18}$O$_2$ experiment (exp. 5), and deposited HCOOH (exp. 6). The patterns are measured for temperatures at 157 K for exps. 2, 4, and 5, and 142 K for exp. 6. \emph{m/z} = 46 and 45 are the masses of the HCOOH$^{+}$ and COOH$^{+}$ ions, respectively.}
\label{fig2}
\end{figure}

\section{Results and discussion}
\label{results}

\subsection{Formation of HCOOH ice by H$_2$CO + H + O$_2$}

\begin{figure}
\includegraphics[width=\columnwidth]{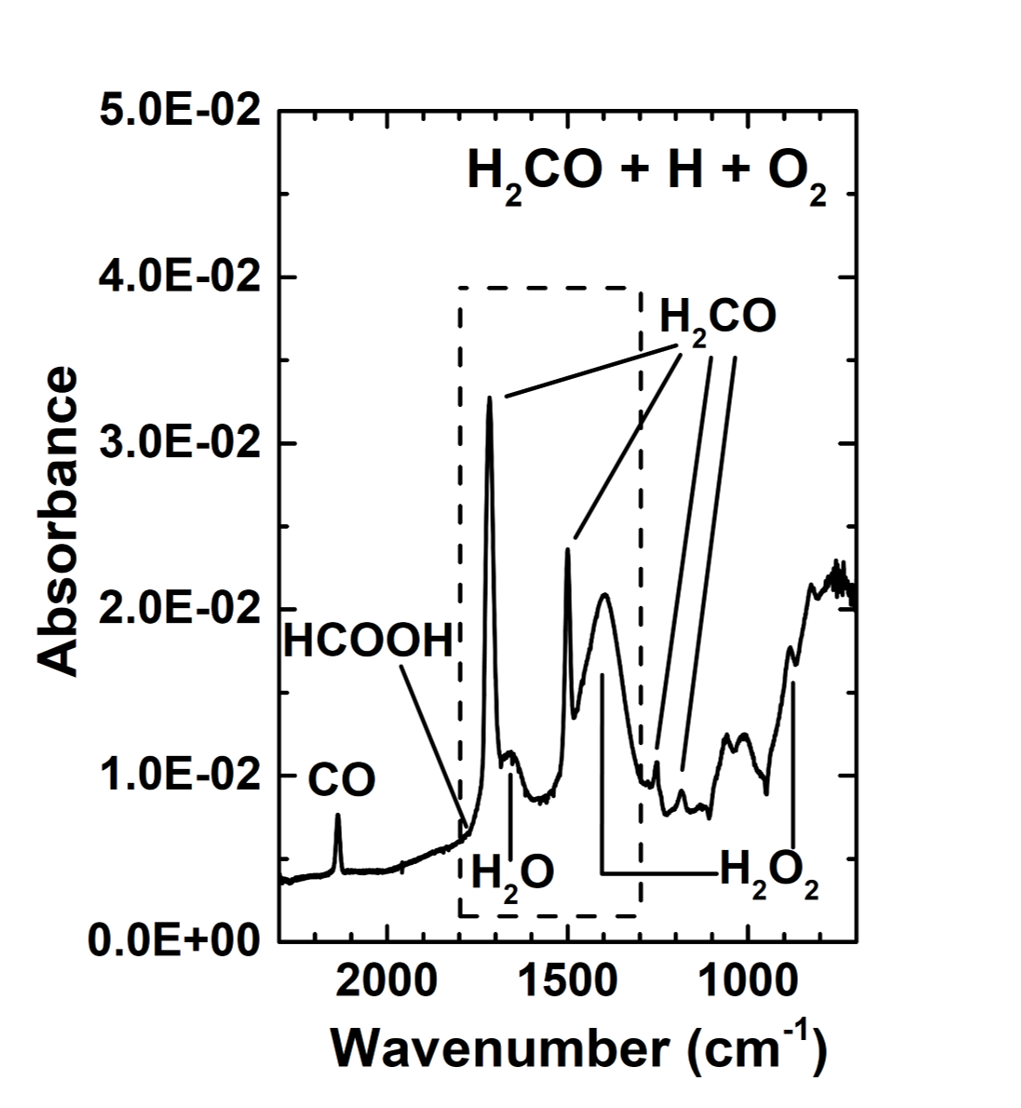}
\caption{RAIR spectrum of H$_2$CO + H + O$_2$ (exp. 1). Spectra are recorded after deposition at 10 K. The dashed-line box is a zoom-in of the spectrum shown in Figure~\ref{fig7}.} 
\label{fig6}
\end{figure}

The top panel of Figure~\ref{fig1} shows the TPD-QMS spectra of newly formed HCOOH obtained after the co-deposition of H$_2$CO + H + O$_2$, as well as pure HCOOH taken after deposition of pure HCOOH. Additional TPD-QMS spectra obtained after the formation of HCOOH isotopologues in correlated isotope-substituted reactions are shown in the bottom panel. Although HCOOH desorbs at 142 K in its pure form (top right panel), the literature shows that HCOOH typically desorbs at a higher temperature when mixed with less volatile compounds \citep{ioppolo2010surface,bennett2010laboratory}, in line with the present experiments. The \emph{m/z} values displayed in the bottom panel of Figure~\ref{fig1} clearly represent the formation of HCOOH involving oxygen atoms that originate from two different molecules, H$_2$CO and O$_2$. If the oxygen atoms of newly formed HCOOH were to solely originate from H$_2$CO, then there would be no desorption signals at 157 K for \emph{m/z} values of 48 and 47 in exp. 4, and for \emph{m/z} values of 49 and 48 in exp. 5. Instead, the signals for \emph{m/z} values of 46 and 45 in exp. 4 and 47 and 46 in exp. 5 would rise, which is not the case. Following this, the \emph{m/z} values presented in the bottom panel of Figure~\ref{fig1} line-up with the formation of a species that contains an oxygen atom from H$_2$CO and an oxygen atom from O$_2$, with \emph{m/z} values that are the same as that of the expected HCOOH isotopologues. Additionally, the desorption peak at 157 K is consistently present among the exps. 2, 4 and 5, and the profiles are nearly identical. This indicates that the desorption peaks should represent the same species and are thus assigned to HCOOH. This assignment is further supported by the QMS fragmentation pattern shown in Figure~\ref{fig2}. HCOOH partially fragments to COOH$^{+}$ upon 70 eV electron impact ionization with a COOH$^{+}$:HCOOH$^{+}$ relative intensity of 78:100 in the pure HCOOH experiment (exp. 6). This relative intensity is similar to the values of 79:100, 79:100, and 82:100 found in the H$_2$CO + H + O$_2$, H$_{2}$CO + H + $^{18}$O$_2$, and H$_{2}$$^{13}$CO + H + $^{18}$O$_2$ experiments, respectively. 

\begin{table*}
\centering
	\caption{List of assigned species in the H$_2$CO + H + O$_2$ (exp. 1) experiment.}
    \begin{threeparttable}
	\label{table2}
	\begin{tabular}{c c c c c} % 10 columns, alignment for each
		\hline
        Peak position & Peak position & Literature values & Molecule & Mode\\
(cm$^{-1}$) & ($\mu$m) & (cm$^{-1}$) \\ % table heading
\hline
880 & 14.64 & 888$^{a}$, 880$^{b}$, 882$^{c}$, 884$^{d}$ & H$_{2}$O$_{2}$ & $\upsilon_3$\\
1184 & 8.446 & 1175$^{e}$, 1175$^{f}$, 1178$^{g}$ & H$_{2}$CO & $\upsilon_6$\\
1251 & 7.994 & 1250$^{e}$, 1253$^{f}$, 1249$^{g}$ & H$_{2}$CO & $\upsilon_5$\\
1396 & 7.163 & 1390$^{a}$, 1368$^{b}$, 1376$^{c}$, 1381$^{d}$& H$_{2}$O$_{2}$ & $\upsilon_{2}$\\
1499 & 6.671 & 1499$^{e}$, 1499$^{f}$, 1499$^{g}$ & H$_{2}$CO & $\upsilon_{3}$\\
1652 & 6.053 & 1650$^{h}$, 1653$^{i}$& H$_{2}$O & $\upsilon_2$\\
1717 & 5.824 & 1718$^{e}$, 1727$^{f}$, 1722$^{g}$& H$_{2}$CO & $\upsilon_2$\\
$\sim$1750 shoulder & $\sim$5.714 & 1690$^{e}$, $\sim$1750$^{j}$, $\sim$1750$^{k}$ & HCOOH & $\upsilon_2$\\ 
2137 & 4.679 & 2141$^{e}$, 2138$^{l}$ & CO & $\upsilon_1$\\
2343 & 4.268 & 2345$^{e}$, 2344$^{l}$ & CO$_{2}$ & $\upsilon_3$\\

\hline
\end{tabular}
\begin{tablenotes}
\item[a]\citet{romanzin2011water}
\item[b]\citet{giguere1959infrared}
\item[c]\citet{lannon1971infrared}
\item[d]\citet{qasim2018formation}
\item[e]\citet{bennett2010laboratory}
\item[f]\citet{chuang2015h}
\item[g]\citet{watanabe2002efficient}
\item[h]\citet{ioppolo2008laboratory}
\item[i]\citet{hodyss2009photochemistry}
\item[j]\citet{ioppolo2010surface}
\item[k]\citet{bisschop2007h}
\item[l]\citet{ioppolo2011surface}
\end{tablenotes}
\end{threeparttable}
\end{table*}

Constraining the formation of HCOOH at 10 K by RAIR data is discussed below. In Figure~\ref{fig6}, a spectrum of H$_2$CO + H + O$_2$ taken after deposition at 10 K is shown. The products formed in the H$_2$CO + H + O$_2$ experiment are listed in Table~\ref{table2} along with the corresponding IR signatures that are labeled in Figure~\ref{fig6}. Note that the two peaks around 1000 cm$^{-1}$ are within the frequency range of the C-O stretch of CH$_3$OH \citep{dawes2016using}. However, these peaks disappear between 195 and 205 K (not shown here), which is a temperature range that is higher than the desorption temperature of CH$_3$OH or any species that are positively identified in this study. Therefore those peaks, although pronounced, are not attributed to a particular species at this time. Yet CH$_3$OH, an expected product, is detected in the TPD-QMS experiments (not shown here). The specific features in Table~\ref{table2} arise from precursor species, intermediates, and reaction products. Some of these species, although formed abundantly in the experiment, are yet to be observed in space. However, certain conditions in the laboratory are vastly different from the conditions in the ISM, and therefore comparison of even relative product abundances from the laboratory must be taken with caution to the relative abundances found in space. For example, H$_2$O$_2$, which is a side product from the H + O$_2$ reaction, has yet to be detected in interstellar ices. It could be efficiently destroyed by a mechanism that does not take place in our experiments, for example. 

\begin{figure*}
\includegraphics[width=\textwidth]{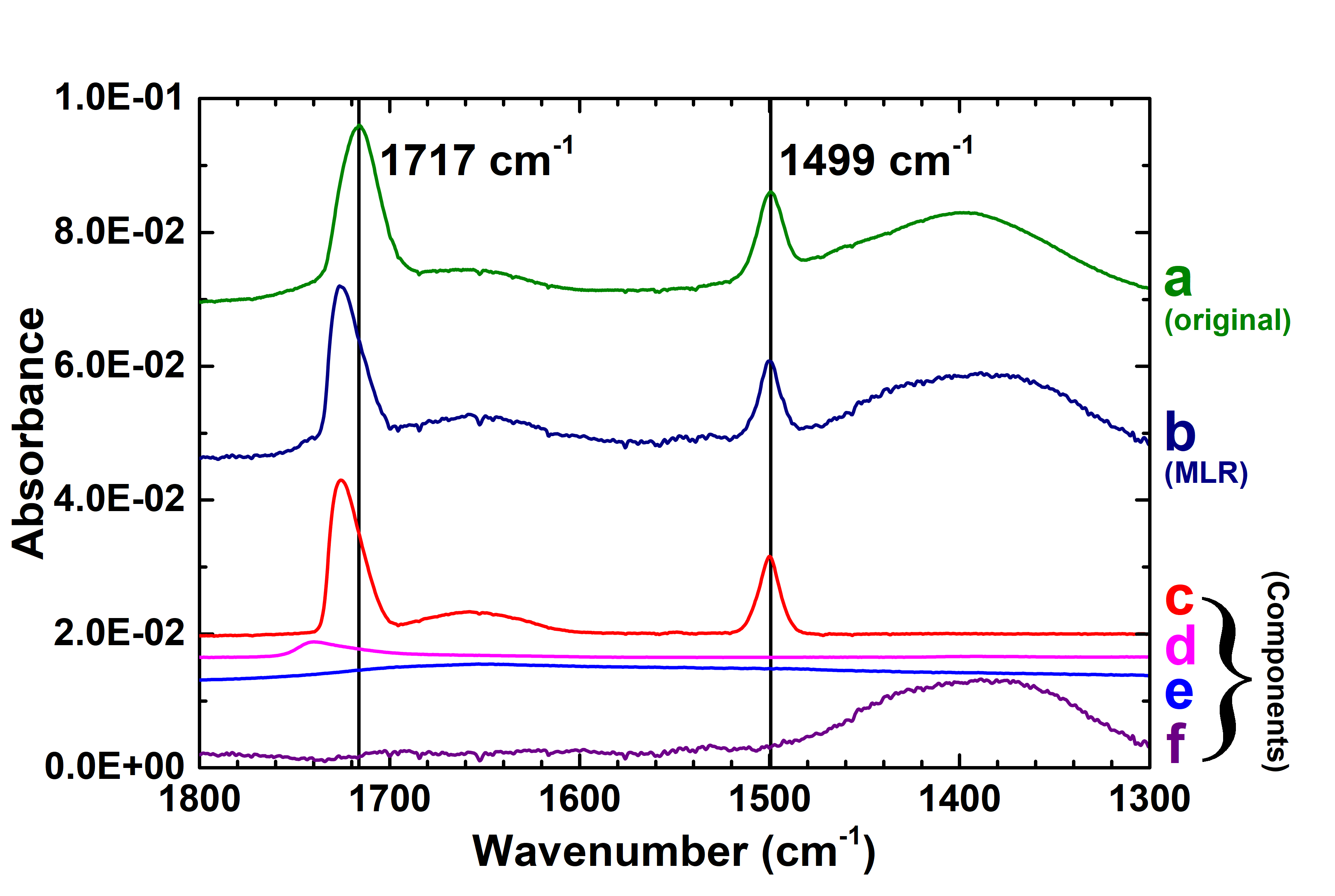}
\caption{RAIR spectra of H$_2$CO + H + O$_2$ (a; exp. 1), the H$_2$CO + H + O$_2$ MLR spectrum (b), and its components multiplied by the coefficients derived from the MLR analysis: H$_2$CO + O$_2$ + H$_2$O (c; exp. 7; 1.2 coefficient), HCOOH (d; exp. 6; 0.3 coefficient), H$_2$O (e; exp. 8; 0.7 coefficient), and H + O$_2$ (f; exp. 9; 9.7 coefficient). Spectra are recorded after deposition at 10 K and are offset for clarity.} 
\label{fig7}
\end{figure*}

Although not straightforward, it is possible to show that HCOOH can be identified spectroscopically. For this, the spectrum is interpreted in a multiple linear regression (MLR) analysis. Figure~\ref{fig7} shows a zoom-in of the spectrum along with spectra of the independent variables. The dependent variable is the original spectrum from Figure~\ref{fig6}. The independent variables represent the expected spectral components of the H$_2$CO + H + O$_2$ experiment. The differences between the components (e.g., HCOOH, H$_2$O, etc.) and the original spectrum (i.e., H$_2$CO + H + O$_2$), as well as the choice of the selected components, are discussed. 

\begin{figure}
\includegraphics[width=\columnwidth]{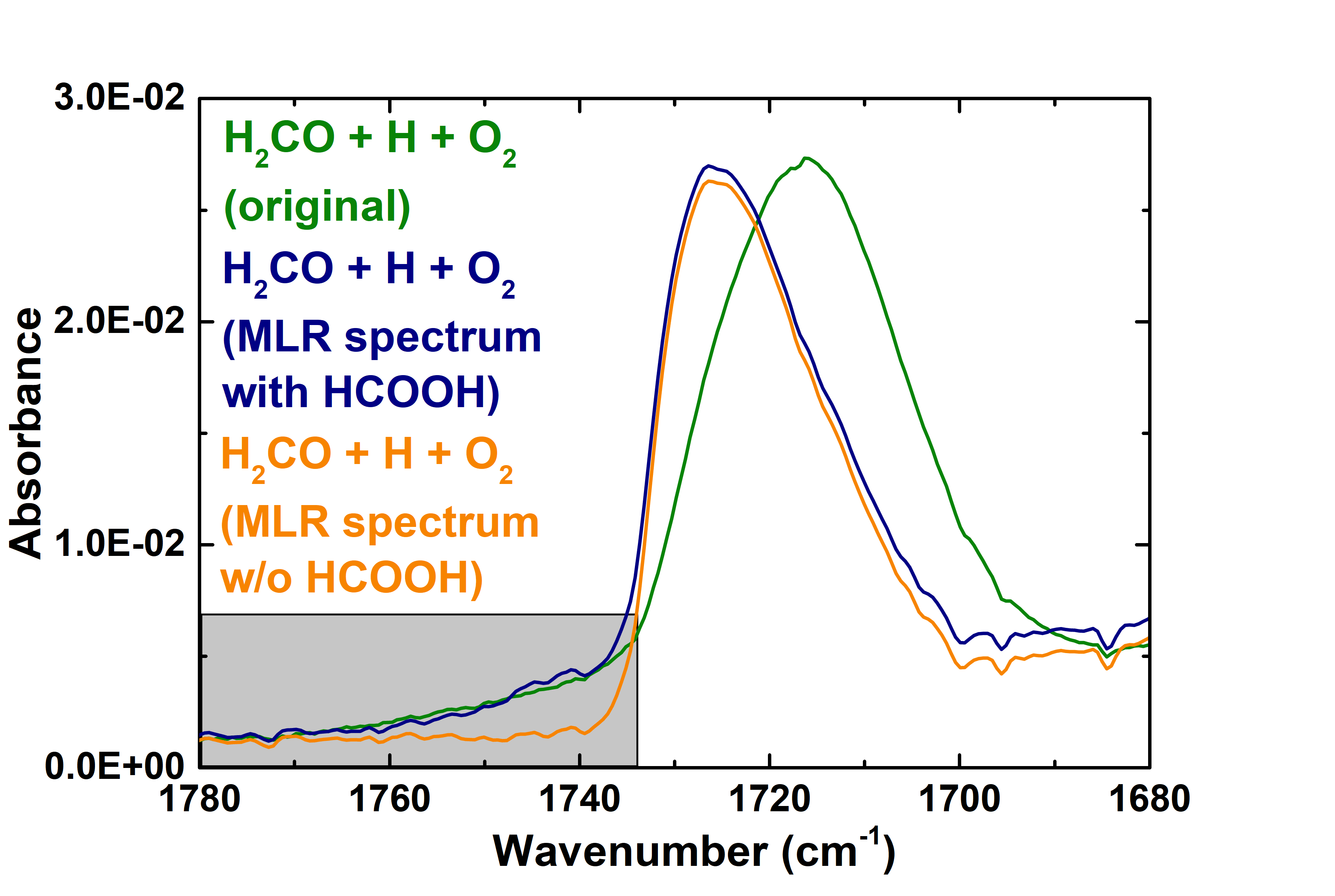}
\caption{RAIR spectra of H$_2$CO + H + O$_2$ (exp. 1) and H$_2$CO + H + O$_2$ MLR spectra with and without HCOOH inclusion (i.e., component d from Figure~\ref{fig7}). The grey box shows the shoulder found in the RAIR spectrum of exp. 1 that is assigned to the C=O stretching mode of HCOOH. Spectra are recorded after deposition at 10 K and are offset for clarity.} 
\label{fig8}
\end{figure}

It is evident in Figure~\ref{fig7} that the H$_2$CO C=O stretching feature at 1717 cm$^{-1}$ in the original spectrum (a) is shifted from that of the MLR spectrum (b), but this is not the case for the C-O stretching feature at 1499 cm$^{-1}$. This indicates that the 1717 cm$^{-1}$ band is more sensitive to the ice mixture ratio and content than the 1499 cm$^{-1}$ band. Therefore, the choice of using H$_2$CO + O$_2$ + H$_2$O (c) versus pure H$_2$CO is to witness how O$_2$ and H$_2$O contribute to broadening and shifting of the C=O peak. The pure H$_2$O component (e) is relatively small, as the majority of H$_2$O is already in the H$_2$CO + O$_2$ + H$_2$O component. The H + O$_2$ component (f) is used to contribute pure H$_2$O$_2$ (i.e., no H$_2$O formed) into the analysis, which is a molecule that is not possible to deposit under our experimental conditions. The contribution of pure H$_2$O$_2$ can probably explain why there is a difference in the profile shapes of the $\sim$1400 cm$^{-1}$ feature between the H + O$_2$ component and the original spectrum. Finally, the contribution of HCOOH (d) can reproduce the $\sim$1750 cm$^{-1}$ shoulder of the original spectrum, as shown in Figure~\ref{fig8}. Note that the HCOOH out-of-phase C=O stretching feature is also sensitive to its surrounding environment \citep{bisschop2007infrared}, and can range from $\sim$1700 cm$^{-1}$ \citep{bisschop2007infrared,bennett2010laboratory} to $\sim$1750 cm$^{-1}$ \citep{bisschop2007h,ioppolo2010surface}. The figure provides zoom-in spectra of the original and MLR spectra from Figure~\ref{fig7}, as well as the MLR spectrum that does not include the HCOOH component. The grey box highlights the shoulder of the C=O stretching feature in the H$_2$CO + H + O$_2$ experiment. When HCOOH is not included, the shoulder disappears. This infers that the shoulder is not solely due to intramolecular broadening effects, and that the inclusion of HCOOH allows to reproduce this feature. This shoulder may also be explained by glycolaldehyde (HCOCH$_2$OH), as the molecule has a signature at $\sim$1750 cm$^{-1}$ and can be formed from the hydrogenation of H$_2$CO \citep{chuang2015h}. Yet, its signal does not appear in the TPD-QMS data and therefore is not expected to appear in the correlating RAIR data. Thus, the RAIR data is fully consistent with the conclusion from the TPD-QMS experiments that HCOOH is formed at 10 K.

\subsection{Formation of CO$_2$ ice by H$_2$CO + H + O$_2$}

\begin{figure}
\includegraphics[width=\columnwidth]{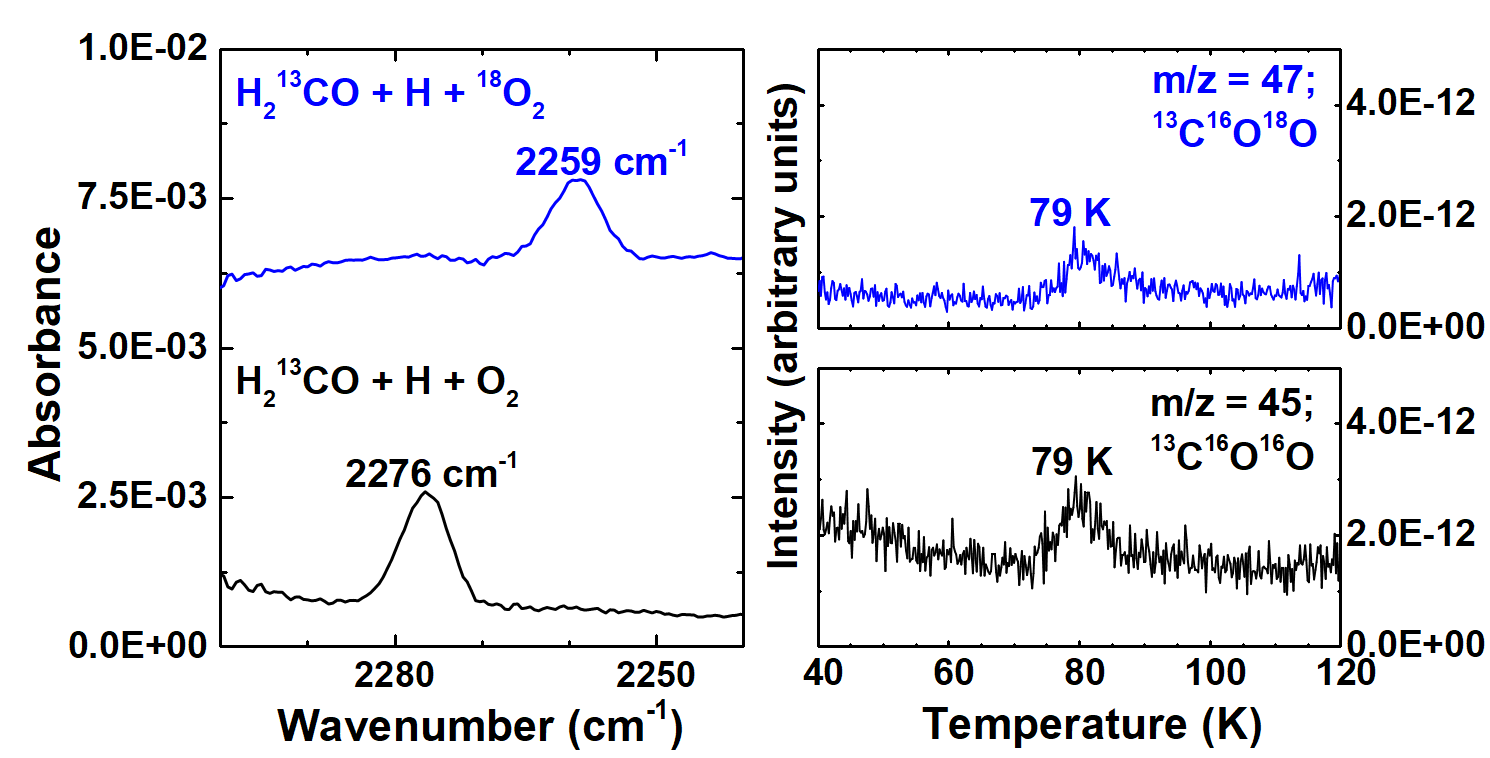}
\caption{(Left) RAIR and (right) TPD-QMS signals that illustrate the characteristic features of $^{13}$COO and $^{13}$CO$^{18}$O found in the H$_{2}$$^{13}$CO + H + O$_2$ (exp. 3) and H$_{2}$$^{13}$CO + H + $^{18}$O$_2$ (exp. 5) experiments, respectively. Both take place at a substrate temperature of 10 K. RAIR spectra are offset for clarity.}
\label{fig5}
\end{figure}

The formation of CO$_2$ ice is confirmed by RAIR and TPD-QMS data presented in Figure~\ref{fig5}. The experiments, H$_{2}$$^{13}$CO + H + O$_2$ and H$_{2}$$^{13}$CO + H + $^{18}$O$_2$, are purposefully displayed in order to not confuse the newly formed CO$_2$ with residual CO$_2$ that is omnipresent as a background pollutant, even under UHV conditions. The RAIR spectra (left panel) clearly show the asymmetric stretching modes of $^{13}$CO$_2$ and $^{13}$CO$^{18}$O at 2276 cm$^{-1}$ and 2259 cm$^{-1}$, respectively \citep{maity2014infrared}. The TPD-QMS data (right panel) also clearly illustrate the desorption of both species at the CO$_2$ desorption temperature of 79 K \citep{fayolle2011laboratory}. We also find that CO$_2$ desorbs at higher temperatures via `volcano' desorption at 150 K (prior to H$_2$O desorption) and co-desorption with H$_2$O$_2$ at 175 K (not shown here). 

\subsection{Pathways to HCOOH and CO$_2$ that are formed in the experiments}

In order to tightly constrain the formation routes of HCOOH and CO$_2$ in our experiments, knowledge of the possible reactions taking place along with the activation barriers and branching ratios are needed. Table~\ref{table3} lists reactions that are expected to occur in the H$_2$CO + H + O$_2$ experiment. Note that the values in the table are from predominantly theoretical studies, as shown in the footnotes. 

\subsubsection{HCOOH formation pathway}
\label{hcoohform}

\begin{figure}
\includegraphics[width=\columnwidth]{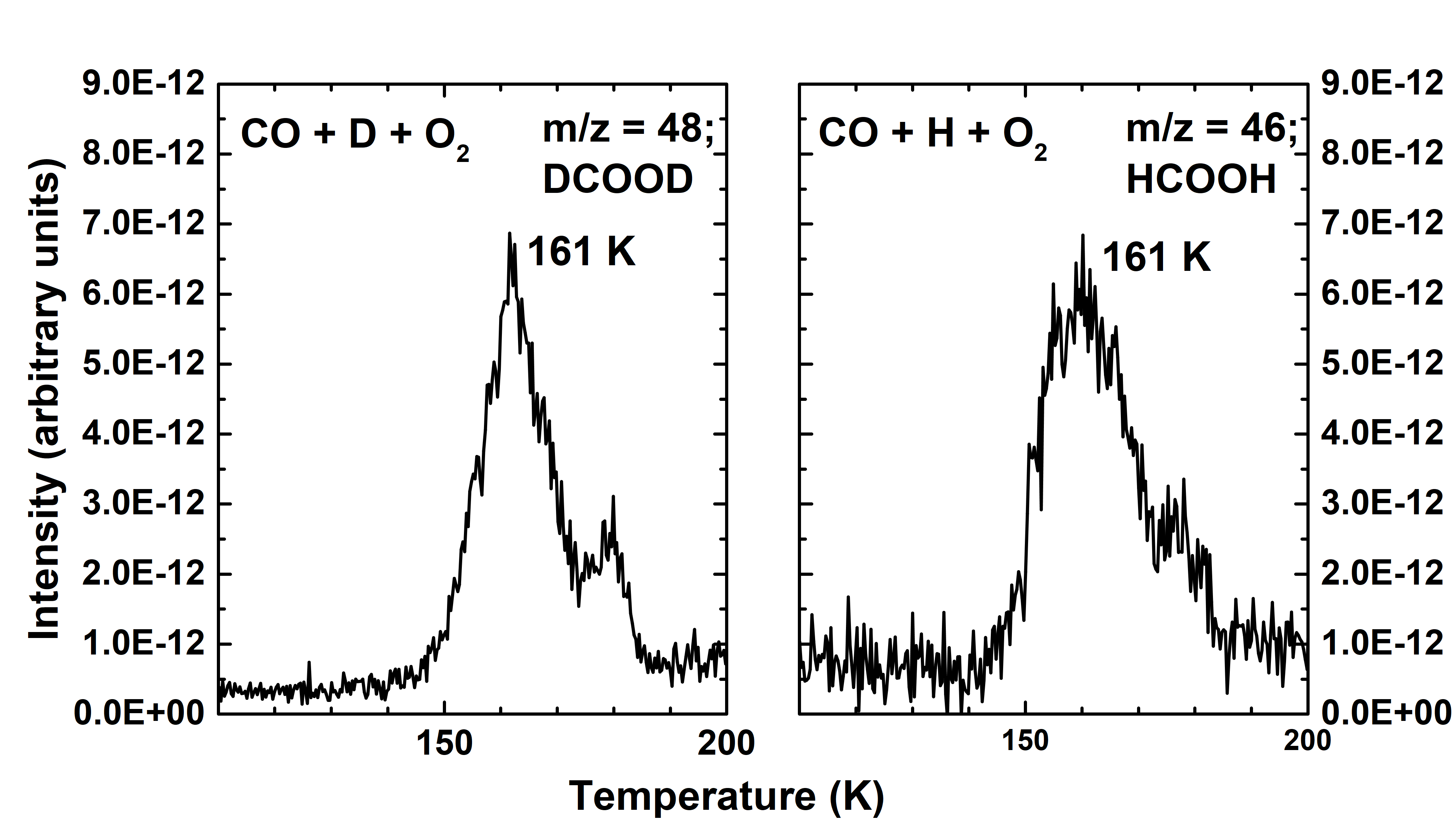}
\caption{TPD-QMS spectra of CO + D + O$_2$ (exp. 10) and CO + H + O$_2$ (exp. 11) recorded after ice growth at 10 K.}
\label{fignew}
\end{figure}

HCOOH can be formed by reactions~\ref{eq01} and/or~\ref{eq1a} in the presented experiments: HCO + OH and/or H + HOCO. The formation of HCO occurs by H-atom addition of CO and H-atom abstraction from H$_2$CO; both which contain activation barriers \citep{andersson2011tunneling,song2017tunneling}. This was demonstrated experimentally by \citet{hidaka2004conversion} and \citet{chuang2015h}. The OH radical is formed by H + HO$_2$, which is a barrierless reaction that occurs early within the H + O$_2$ reaction chain \citep{lamberts2013water}. It is also formed by the H + H$_2$O$_2$ reaction \citep{lamberts2017influence}, which however involves barriers. The activation energy of HOCO formed by OH + CO is calculated to be nearly zero \citep{nguyen2012reaction,masunov2016chemical,tachikawa2016effects}, and since H + HOCO is barrierless, one may expect that H + HOCO dominates HCOOH formation. Yet, the pathway also requires two extra steps (i.e., H-abstraction from HCO and formation of HOCO) in comparison to the HCO + OH route. To determine the relevance of the two pathways, experimental and theoretical data are combined and the outcomes are discussed below.  

\begin{table*}
\centering
	\caption{List of possible reactions taking place in the H$_2$CO + H + O$_2$ experiment.}
    \begin{threeparttable}
	\label{table3}
	\begin{tabular}{c c c c c c} % 10 columns, alignment for each
		\hline
        Reaction & Product(s) & Branching ratio & Activation energy & Rate constant & Reference\\
& & (\%) & (kJ/mol) & (s$^{-1}$) \\ % table heading
\hline
H + H$_2$CO $\rightarrow$ & CH$_3$O & - & 16-18 & $6 \times 10^{5}$ - $2 \times 10^{6}$ & a \\
 & CH$_2$OH & - & 43-47 & $4 \times 10^{1}$ - $9 \times 10^{1}$ & a\\
& H$_2$ + HCO & - & 21-25 & $4 \times 10^{5}$ - $1 \times 10^{6}$ & a\\
\\
H + HCO $\rightarrow$ & H$_2$CO & 50 & 0 & -& zeroth-order approximation$^{\spadesuit}$\\
 & H$_2$ + CO & 50 & 0 & - & zeroth-order approximation\\
 \\
H + CO $\rightarrow$ & HCO & - & 12.4$^b$ + 3$^c$ & $2 \times 10^{5}$ & b, c\\
\\
H + O$_2$ $\rightarrow$ & HO$_2$ & 100 & $\sim$0 & - & d\\
\\
H + HO$_2$ $\rightarrow$ & 2 OH & 50 & 0 & - & e, zeroth-order approximation\\
 & H$_2$O$_2$ & 50 & 0 & - & e, zeroth-order approximation\\
 \\
2 OH $\rightarrow$ & H$_2$O$_2$ & 90 & 0 & - & d\\
 & H$_2$O + O & 10 & 0 & - & d\\
 \\
H + H$_2$O$_2$ $\rightarrow$ & H$_2$O + OH & - & 21-27 & $2 \times 10^{3}$ - $1 \times 10^{6}$ & f\\
 & H$_2$ + HO$_2$ & - & 39 & $\diamond$ & g\\
 \\
OH + HCO $\rightarrow$ & HCOOH & 50 & 0 & - & zeroth-order approximation\\
 & H$_2$O + CO & 50 & 0 & - & zeroth-order approximation\\
 \\
OH + H$_2$CO $\rightarrow$ & H$_2$O + HCO & - & 2.64 & $\diamond$ & h\\
\\
OH + CO $\rightarrow$ & H + CO$_2$ & - & * & * & i, j, k\\
 & HOCO & $\sim$100 & $\sim$0 & - & i, j, k\\
 \\
H + HOCO $\rightarrow$ & H$_2$ + CO$_2$ & 50 & 0 & - & zeroth-order approximation\\
 & HCOOH & 50 & 0 & - & zeroth-order approximation\\
 \\
OH + H $\rightarrow$ & H$_2$O & 100 & 0 & - & -\\
\\
OH + H$_2$ $\rightarrow$ & H$_2$O + H & - & 22.4-24.3 & $2 \times 10^{5}$ - $5 \times 10^{5}$ & l\\
\\
H + CH$_3$O $\rightarrow$ & CH$_3$OH & 50 & 0 & - & zeroth-order approximation\\
 & H$_2$CO + H$_2$ & 50 & 0 & - & zeroth-order approximation\\
\hline
\end{tabular}
\begin{tablenotes}
\item[a]\citet{song2017tunneling}
\item[b]\citet{andersson2011tunneling}
\item[c]\citet{alvarez2018hydrogen}; zero-point energy contribution
\item[d]\citet{lamberts2013water}
\item[e]\citet{lamberts2014formation}
\item[f]\citet{lamberts2017influence}
\item[g]\citet{lamberts2016quantum}\\
\item[h]\citet{zanchet2018full}
\item[i]\citet{nguyen2012reaction}
\item[j]\citet{masunov2016chemical}
\item[k]\citet{tachikawa2016effects}
\item[l]\citet{meisner2017atom}

(-) indicates values that are not the defining parameters.\\ 
(*) indicates multi-step reaction; values cannot be trivially obtained.\\ 
($\diamond$) indicates that unimolecular rate constants are not available.\\
($^{\spadesuit}$) indicates first educated guess. 

\end{tablenotes}
\end{threeparttable}
\end{table*}

The formation of CO ice is shown in Figure~\ref{fig6} at 2137 cm$^{-1}$. This occurs by the hydrogen abstraction from H$_2$CO to produce HCO, and subsequently the hydrogen abstraction from HCO to produce CO. This HCO radical can react with a nearby OH radical barrierlessly to form HCOOH. Next, the contribution from the H + HOCO pathway is discussed. The HOCO intermediate is not observed in the RAIR spectra, as it was in other studies \citep{oba2010experimental,ioppolo2010surface}. However, HOCO is an unstable species, and can be stabilized depending on the polarity of the ice matrix \citep{ioppolo2011surface} and whether HOCO is embedded in the ice \citep{arasa2013molecular}. Therefore, HOCO can still be relevant to HCOOH formation in the H$_2$CO + H + O$_2$ experiment, even though it is undetected in our RAIR data. According to \citet{ioppolo2010surface}, in a CO-rich ice, H + HOCO is found to be the dominant pathway to HCOOH formation. Thus, with the sight of CO in our RAIR data as shown in Figure~\ref{fig6}, it cannot be excluded that H + HOCO is a contributing route for HCOOH formation. As the dominance of the H + HOCO pathway was determined by one weak transition of HOCO in \citet{ioppolo2010surface}, we performed isotope experiments with CO to further confirm this.

Figure~\ref{fignew} shows TPD-QMS spectra of \emph{m/z} = 48 (representative of DCOOD) from the CO + D + O$_2$ experiment and \emph{m/z} = 46 (representative of HCOOH) from the CO + H + O$_2$ experiment. In essence, the relative abundance of DCOOD to HCOOH shows which pathway dominates formic acid formation starting from a CO-rich ice. This can be explained by discussion of the following reactions:

\begin{align}
\label{eq7}
\mathrm{OD + CO} & \rightarrow \mathrm{DOCO}\overset{\text{+D}}{\to} \mathrm{DCOOD}\\
\label{eq8}
\mathrm{OH + CO} & \rightarrow \mathrm{HOCO}\overset{\text{+H}}{\to} \mathrm{HCOOH}
\end{align}

OD/OH + CO is essentially barrierless, and D/H + DOCO/HOCO is barrierless. Thus, the reaction rates of reactions~\ref{eq7} and ~\ref{eq8} are similar. However, for the reactions below:  

\begin{align}
\label{eq9}
\mathrm{D + CO} & \rightarrow \mathrm{DCO}\overset{\text{+OD}}{\to} \mathrm{DCOOD}\\
\label{eq10}
\mathrm{H + CO} & \rightarrow \mathrm{HCO}\overset{\text{+OH}}{\to} \mathrm{HCOOH} 
\end{align}

D + CO is more than two orders of magnitude slower than H + CO \citep{andersson2011tunneling}. Thus, if the abundance of HCOOH is much greater than the abundance of DCOOD, then HCO + OH would be considered the more dominant pathway. 

The TPD-QMS results from Figure~\ref{fignew} show that the integrated areas of DCOOD and HCOOH are essentially the same (i.e., they are not different by orders of magnitude). Therefore, it cannot be claimed from the presented results that HCO + OH is dominating in the CO + H + O$_2$ experiment. Rather, H + HOCO dominates a CO-rich ice, which is in agreement with the findings from \citet{ioppolo2010surface}. However, which pathway contributes more or less to the formation of HCOOH in the H$_2$CO + H + O$_2$ experiment cannot be extracted here. We restrict our conclusion to the finding that both pathways occur and contribute to the formation of HCOOH. This is for the conditions investigated in our setup. To extrapolate these to interstellar ices, it is important to use astrochemical simulations in order to compare the relative contributions of each of the suggested reaction pathways under dark cloud conditions.

\subsubsection{CO$_2$ formation pathway}

The formation of CO$_2$ is also via two pathways, as shown in Table~\ref{table3}. Note that OH + HOCO is not considered, as OH + HOCO should yield carbonic acid (H$_2$CO$_3$) \citep{oba2010formation,ioppolo2010surface}, in addition to CO$_2$ \citep{yu2005direct,francisco2010hoco}, and H$_2$CO$_3$ is not detected in our experiments. OH + CO $\rightarrow$ HOCO has an activation barrier of nearly zero, whereas OH + CO $\rightarrow$ CO$_2$ + H has barriers, making the HOCO product a more likely outcome. HOCO can then react with an H-atom to form H$_2$ + CO$_2$ and HCOOH without a barrier. In section~\ref{hcoohform}, we discuss that H + HOCO must be a route to HCOOH formation in the H$_2$CO + H + O$_2$ experiment. According to Table~\ref{table3}, CO$_2$ is then also expected to be formed alongside HCOOH via H + HOCO. By combination of the experimental results in section~\ref{hcoohform} with the theoretical values from Table~\ref{table3}, we propose that H + HOCO is the dominating pathway to CO$_2$ formation in our experiments. Our experimental condition of co-deposition is also suitable for forming the HOCO complex, as the HOCO complex survives when it is able to dissipate its energy throughout the ice \citep{arasa2013molecular}. This excess energy is lost to the ice matrix within picoseconds \citep{arasa2010molecular,fredon2017energy}, making it less likely that there is enough time for OH + CO $\rightarrow$ CO$_2$ + H to be attempted. Note that the H$_2$CO + O pathway is unlikely in the presented experiments, since H-atoms will actively compete with H$_2$CO to react with O-atoms. CO + O is also not listed, as the barrier of this reaction \citep{roser2001formation} makes it a relatively minor reaction channel \citep{goumans2010tunnelling,ioppolo2013surfreside2}. The reaction of H + CO$_2$ has a very high barrier of > $\sim$15,000 K \citep{bisschop2007h, mccarthy2016isotopic}, which is too high to allow for tunneling to speed up the reaction considerably.  

Since CO$_2$ is predominantly formed by H + HOCO, the H + HOCO branching ratio can be used, in combination with the HCOOH:CO$_2$ ice abundance ratio of around 1.8:1, to determine the contributions of HCO + OH and H + HOCO to HCOOH formation in our experiments. Yet, as shown in Table~\ref{table3}, the H + HOCO branching ratio is based off an approximation. Thus, a more well-defined branching ratio is desired in order to perform such a quantitative analysis.  

\subsubsection{Surface reaction mechanism}

The dominating reaction mechanism in our experiments is discussed. Typically, interstellar ice analogues are formed by three mechanisms: Langmuir-Hinshelwood (L-H), Eley-Rideal (E-R), and hot-atom (H-A) \citep{linnartz2015atom, he2017mechanism}. At the low temperature of 10 K, H-atoms have a high enough sticking coefficient to diffuse through the ice and react with other ice constituents. However, as the surface temperature increases, the H-atom residence time dramatically decreases. This phenomenon has been demonstrated in a number of laboratory works \citep{watanabe2002efficient,watanabe2003dependence,cuppen2007simulation,fuchs2009hydrogenation,chuang2015h,qasim2018formation}, where hydrogenation product abundances significantly decrease as the substrate temperature rises beneath the initial desorption point of the reactant molecule(s). These observations show that the product abundance is governed by the substrate temperature, which is in favor of the L-H mechanism as the dominating mechanism in these studies. This then also concludes that the majority of H-atoms are thermally equilibrated to the 10 K surface prior to reaction.

We also find that tunneling is essential to product formation in this study. As shown in Table~\ref{table3}, H + H$_2$CO has high activation barriers that range from 16-47 kJ/mol. However it is shown here, and also in \citet{chuang2015h}, that the H-induced abstraction of two H-atoms from H$_2$CO to form CO occurs. With such high barrier values, the H-atom would not have sufficient energy to hop over the barrier, and therefore must undergo tunneling. The importance of tunneling in these type of reactions is discussed in more detail in \citet{lamberts2017importance}.           

\section{Astrophysical implications}
\label{astro}

\begin{figure}
\includegraphics[width=\columnwidth]{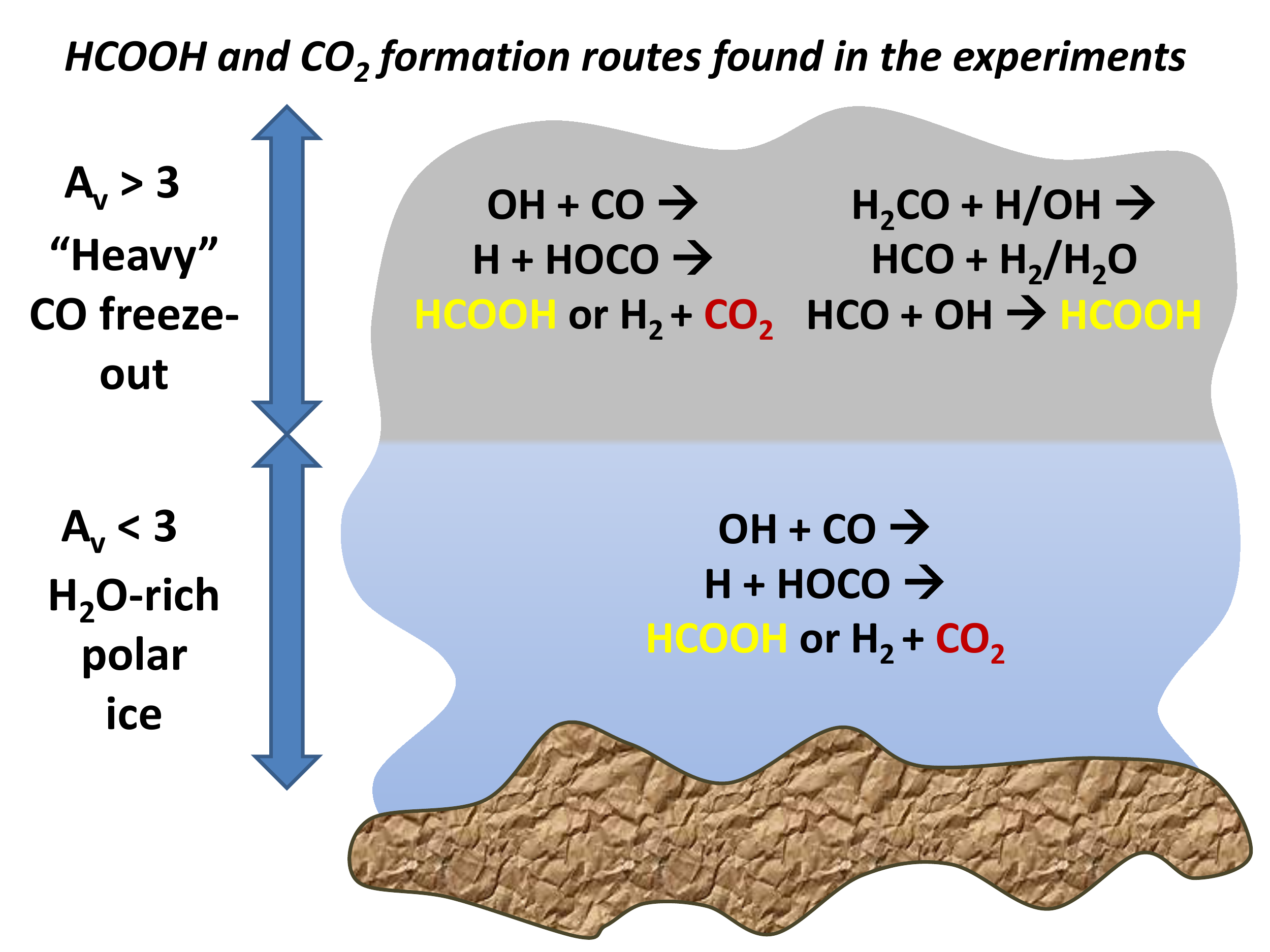}
\caption{Illustration of the formation pathways of HCOOH (highlighted in yellow) and CO$_2$ (highlighted in red) ice in the H$_2$O-rich and heavy CO freeze-out stages, as proposed only from the findings from this study (i.e., not all possible reactions are included). Note that an extra pathway to HCOOH formation is found in the heavy CO freeze-out stage due to the presence of H$_2$CO.}
\label{cartoon}
\end{figure}

Grain surface chemistry is a strong function of cloud depth, as the increasing density into the cloud enhances the gas-grain interaction. With the interstellar radiation field decreasing at larger extinctions (A$_V$) by dust particles, photo-induced processes become less relevant in comparison to `non-energetic' processes, such as hydrogenation \citep{chuang2017production}. Thus, we expect the `non-energetic' solid-state HCOOH formation route(s) investigated in our experiments, as well as their efficiencies, to depend strongly on cloud depth. The processes involved are complex and require detailed models in order to quantify their contribution to HCOOH and CO$_2$ ice formation in the CO freeze-out stage. However, such processes are roughly expected to take place as follows.

Following \citet{cuppen2009microscopic} and \citet{tielens1982model}, the gas-phase CO:H and O:H ratios are critical parameters in grain surface chemistry. Initially, at cloud depths corresponding to A$_V$ values of less than a few magnitudes, the O-rich gas rapidly becomes hydrogenated on grain surfaces. Subsequently, a polar ice that is rich in H$_2$O is formed, and the abundantly formed intermediate product, OH, will also react with CO. Our study points to the conclusion that in this environment, this reaction is more likely to form the intermediate HOCO than directly CO$_2$. Competition by the reaction of CO with H is low at this stage because of the high abundance of OH on the grains and the relatively high gas-phase CO:H ratio, as CO is not yet frozen out. With the presence of HOCO, both CO$_2$ and HCOOH are then expected to be formed. Indeed, the observed CO$_2$ abundance is very high ($\sim$20\% of H$_2$O \citet{bergin2005spitzer}). The identification of HCOOH at low extinctions is tentative at best (< 5\% \citep{boogert2011ice}), because its strongest mode ($\sim$6.0 $\mu$m) overlaps with that of the H$_2$O bending mode.

At visual extinctions above $\sim$3 magnitude, the gas-phase CO:H ratio decreases as more CO freezes out. Also, most of the O is locked up in H$_2$O, reducing the OH abundance on the grain surface. This promotes CO hydrogenation (H + CO). The formation of HOCO becomes less relevant, as the barrierless HCO + OH reaction will contribute to the formation of HCOOH. The formation of CO$_2$ is then expected to be less important, although our experiments show that CO$_2$ is formed in the ice. At ever increasing extinctions and densities into the cloud, HCO preferably hydrogenates to H$_2$CO. H$_2$CO ice has indeed likely been detected towards Young Stellar Objects at the 2-7\% level with respect to H$_2$O ice \citep{keane2001ice, boogert2015observations}. The observations are currently of insufficient quality to determine if H$_2$CO is present in the polar (H$_2$O-rich) or apolar (CO-rich) ice phases, but the latter is most likely \citep{cuppen2009microscopic}. H-abstraction reactions will maintain an HCO reservoir and thus the HCO + OH route to HCOOH will continue. From the presented experiments, it is also observed that H + HOCO takes place in the H$_2$CO + H + O$_2$ experiment to form HCOOH. With two formation routes involving H$_2$CO to form HCOOH, it is expected that the HCOOH:CO$_2$ ratio is highest at A$_V$ > 3. Nevertheless, due to the low OH abundance, the absolute HCOOH abundance (with respect to dust) in the CO-rich layer is probably still less than that at lower extinctions in the H$_2$O-rich layer. An illustration of the relevance of these reactions to the H$_2$O-rich and heavy CO freeze-out stages are schematically shown in Figure~\ref{cartoon}. Note that the figure only includes the formation mechanisms found in this work and does not provide an overview of all expected/studied mechanisms. Observations have shown that at A$_V$ > 9, where the "catastrophic" CO freeze-out takes place, CH$_3$OH becomes dominant \citep{boogert2011ice}, and this is expected to be a less favorable environment for HCOOH formation due to the lack of HCO radicals. An important caveat is that the CH$_3$OH formation threshold of 9 mag is very uncertain. In fact, in some molecular cores, no CH$_3$OH ice is observed well above
this threshold \citep{boogert2011ice}. Following the discussion above, this could enhance HCOOH abundances.

 A search for HCOOH at extinctions across the entire extinction range is warranted. The best band to detect HCOOH is at 7.25 $\mu$m \citep{schutte1999weak} and has been seen only toward a few YSO envelopes \citep{oberg2011spitzer, boogert2015observations}. In a few years, however sensitive searches in prestellar sightlines across a wide A$_V$ range will be possible with the James Webb Space Telescope (JWST). Studies of the absorption band profiles (peak positions and widths) for a range of sightlines are needed to secure the identification of HCOOH in H$_2$O and CO-rich sightlines.

\section{Conclusions}
\label{conc}

The primary findings of this study are listed below:

\begin{itemize}

\item The hydrogenation of an H$_2$CO:O$_2$ ice mixture to study the H$_2$CO + OH reaction at 10 K adds another `non-energetic' formation route to the solid-state formation of HCOOH and CO$_2$ in cold and dark interstellar clouds. Astrochemical modeling is desired to know to what extent the new reactions added here contribute.\\

\item The formation of HCOOH in the H$_2$CO + H + O$_2$ experiment occurs by both, H + HOCO and HCO + OH. The exact value of their relative contributions can be determined once a more well-defined branching ratio of H + HOCO becomes available.\\ 

\item The formation of HCOOH in the CO + H + O$_2$ experiment occurs predominantly by H + HOCO, as shown here and in \citet{ioppolo2010surface}.\\

\item The formation of CO$_2$ in both, H$_2$CO + H + O$_2$ and CO + H + O$_2$ experiments, is predominantly through H + HOCO rather than OH + CO for the conditions studied here.\\ 

\item A search for HCOOH ice in the ISM is expected to be promising at extinctions where HCO and H$_2$CO are formed, but before H$_2$CO is sufficiently hydrogenated to form CH$_3$OH, which albeit has an uncertain formation threshold of 9 mag.   

\end{itemize}

\begin{acknowledgements}
      This research would not have been possible without the financial support from the Dutch Astrochemistry Network II (DANII). Further support includes a VICI grant of NWO (the Netherlands Organization for Scientific Research) and funding by NOVA (the Netherlands Research School for Astronomy). D.Q. thanks Marc van Hemert for insightful discussions. T.L. is supported by the NWO via a VENI fellowship (722.017.00). G.F. recognizes the financial support from the European Union's Horizon 2020 research and innovation programme under the Marie Sklodowska-Curie grant agreement n. 664931. S.I. recognizes the Royal Society for financial support and the Holland Research School for Molecular Chemistry (HRSMC) for a travel grant.
\end{acknowledgements}
\bibliography{hcooh}

\begin{thebibliography}{122}
\expandafter\ifx\csname natexlab\endcsname\relax\def\natexlab#1{#1}\fi

\bibitem[{{\'A}lvarez-Barcia {et~al.}(2018){\'A}lvarez-Barcia, Russ,
  K{\"a}stner, \& Lamberts}]{alvarez2018hydrogen}
{\'A}lvarez-Barcia, S., Russ, P., K{\"a}stner, J., \& Lamberts, T. 2018, MNRAS,
  479, 2007

\bibitem[{Andersson {et~al.}(2011)Andersson, Goumans, \&
  Arnaldsson}]{andersson2011tunneling}
Andersson, S., Goumans, T., \& Arnaldsson, A. 2011, Chem. Phys. Lett., 513, 31

\bibitem[{Andrade {et~al.}(2013)Andrade, de~Barros, Pilling, Domaracka,
  Rothard, Boduch, \& da~Silveira}]{andrade2013chemical}
Andrade, D.~P., de~Barros, A.~L., Pilling, S., {et~al.} 2013, MNRAS, 430, 787

\bibitem[{Arasa {et~al.}(2010)Arasa, Andersson, Cuppen, van Dishoeck, \&
  Kroes}]{arasa2010molecular}
Arasa, C., Andersson, S., Cuppen, H., van Dishoeck, E., \& Kroes, G.-J. 2010,
  J. Chem. Phys., 132, 184510

\bibitem[{Arasa {et~al.}(2013)Arasa, van Hemert, van Dishoeck, \&
  Kroes}]{arasa2013molecular}
Arasa, C., van Hemert, M.~C., van Dishoeck, E.~F., \& Kroes, G.-J. 2013, J.
  Phys. Chem. A, 117, 7064

\bibitem[{Bennett {et~al.}(2010)Bennett, Hama, Kim, Kawasaki, \&
  Kaiser}]{bennett2010laboratory}
Bennett, C.~J., Hama, T., Kim, Y.~S., Kawasaki, M., \& Kaiser, R.~I. 2010, ApJ,
  727, 27

\bibitem[{Bennett {et~al.}(2009{\natexlab{a}})Bennett, Jamieson, \&
  Kaiser}]{bennett2009experimentalCO}
Bennett, C.~J., Jamieson, C.~S., \& Kaiser, R.~I. 2009{\natexlab{a}}, ApJS,
  182, 1

\bibitem[{Bennett {et~al.}(2009{\natexlab{b}})Bennett, Jamieson, \&
  Kaiser}]{bennett2009mechanistical}
Bennett, C.~J., Jamieson, C.~S., \& Kaiser, R.~I. 2009{\natexlab{b}}, Phys.
  Chem. Chem. Phys., 11, 4210

\bibitem[{Bergantini {et~al.}(2013)Bergantini, Pilling, Rothard, Boduch, \&
  Andrade}]{bergantini2013processing}
Bergantini, A., Pilling, S., Rothard, H., Boduch, P., \& Andrade, D. 2013,
  MNRAS, 437, 2720

\bibitem[{Bergin {et~al.}(2005)Bergin, Melnick, Gerakines, Neufeld, \&
  Whittet}]{bergin2005spitzer}
Bergin, E.~A., Melnick, G.~J., Gerakines, P.~A., Neufeld, D.~A., \& Whittet,
  D.~C. 2005, ApJL, 627, L33

\bibitem[{Bisschop {et~al.}(2007{\natexlab{a}})Bisschop, Fuchs, Boogert,
  Van~Dishoeck, \& Linnartz}]{bisschop2007infrared}
Bisschop, S., Fuchs, G., Boogert, A., Van~Dishoeck, E., \& Linnartz, H.
  2007{\natexlab{a}}, A\&A, 470, 749

\bibitem[{Bisschop {et~al.}(2007{\natexlab{b}})Bisschop, Fuchs, Van~Dishoeck,
  \& Linnartz}]{bisschop2007h}
Bisschop, S., Fuchs, G., Van~Dishoeck, E., \& Linnartz, H. 2007{\natexlab{b}},
  A\&A, 474, 1061

\bibitem[{Boogert {et~al.}(2011)Boogert, Huard, Cook, Chiar, Knez, Decin,
  Blake, Tielens, \& Van~Dishoeck}]{boogert2011ice}
Boogert, A., Huard, T., Cook, A., {et~al.} 2011, ApJ, 729, 92

\bibitem[{Boogert {et~al.}(2015)Boogert, Gerakines, \&
  Whittet}]{boogert2015observations}
Boogert, A.~A., Gerakines, P.~A., \& Whittet, D.~C. 2015, ARA\&A, 53, 541

\bibitem[{Boonman {et~al.}(2003)Boonman, Van~Dishoeck, Lahuis, \&
  Doty}]{boonman2003gas}
Boonman, A., Van~Dishoeck, E., Lahuis, F.~v., \& Doty, S. 2003, A\&A, 399, 1063

\bibitem[{Bottinelli {et~al.}(2007)Bottinelli, Ceccarelli, Williams, \&
  Lefloch}]{bottinelli2007hot}
Bottinelli, S., Ceccarelli, C., Williams, J.~P., \& Lefloch, B. 2007, A\&A,
  463, 601

\bibitem[{Bouilloud {et~al.}(2015)Bouilloud, Fray, B{\'e}nilan, Cottin, Gazeau,
  \& Jolly}]{bouilloud2015bibliographic}
Bouilloud, M., Fray, N., B{\'e}nilan, Y., {et~al.} 2015, MNRAS, 451, 2145

\bibitem[{Butscher {et~al.}(2016)Butscher, Duvernay, Danger, \&
  Chiavassa}]{butscher2016radical}
Butscher, T., Duvernay, F., Danger, G., \& Chiavassa, T. 2016, A\&A, 593, A60

\bibitem[{Chuang(2018, Univ. Leiden)}]{chuang2018formation}
Chuang, K. 2018, Univ. Leiden, PhD thesis

\bibitem[{Chuang {et~al.}(2016)Chuang, Fedoseev, Ioppolo,
  {et~al.}}]{chuang2015h}
Chuang, K.-J., Fedoseev, G., Ioppolo, S., {et~al.} 2016, MNRAS, 455, 1702

\bibitem[{Chuang {et~al.}(2017)Chuang, Fedoseev, Qasim, Ioppolo, van Dishoeck,
  \& Linnartz}]{chuang2017production}
Chuang, K.-J., Fedoseev, G., Qasim, D., {et~al.} 2017, MNRAS, 467, 2552

\bibitem[{Chuang {et~al.}(2018)Chuang, Fedoseev, Qasim, Ioppolo, van Dishoeck,
  \& Linnartz}]{chuang2018reactive}
Chuang, K.-J., Fedoseev, G., Qasim, D., {et~al.} 2018, ApJ, 853, 102

\bibitem[{Cook {et~al.}(2011)Cook, Whittet, Shenoy, Gerakines, White, \&
  Chiar}]{cook2011thermal}
Cook, A., Whittet, D., Shenoy, S., {et~al.} 2011, ApJ, 730, 124

\bibitem[{Cottin {et~al.}(2003)Cottin, Moore, \&
  B{\'e}nilan}]{cottin2003photodestruction}
Cottin, H., Moore, M.~H., \& B{\'e}nilan, Y. 2003, ApJ, 590, 874

\bibitem[{Cuppen \& Herbst(2007)}]{cuppen2007simulation}
Cuppen, H. \& Herbst, E. 2007, ApJ, 668, 294

\bibitem[{Cuppen {et~al.}(2010)Cuppen, Ioppolo, Romanzin, \&
  Linnartz}]{cuppen2010water}
Cuppen, H., Ioppolo, S., Romanzin, C., \& Linnartz, H. 2010, Phys. Chem. Chem.
  Phys., 12, 12077

\bibitem[{Cuppen {et~al.}(2009)Cuppen, Van~Dishoeck, Herbst, \&
  Tielens}]{cuppen2009microscopic}
Cuppen, H., Van~Dishoeck, E., Herbst, E., \& Tielens, A. 2009, A\&A, 508, 275

\bibitem[{Dartois {et~al.}(1998)Dartois, d'Hendecourt, Boulanger, Jourdain~de
  Muizon, Breitfellner, Puget, \& Habing}]{dartois1998molecular}
Dartois, E., d'Hendecourt, L., Boulanger, F., {et~al.} 1998, A\&A, 331, 651

\bibitem[{Dawes {et~al.}(2016)Dawes, Mason, \& Fraser}]{dawes2016using}
Dawes, A., Mason, N.~J., \& Fraser, H.~J. 2016, Phys. Chem. Chem. Phys., 18,
  1245

\bibitem[{De~Graauw {et~al.}(1996)De~Graauw, Whittet, Gerakines, Bauer,
  Beintema, Boogert, Boxhoorn, Chiar, Ehrenfreund, Feuchtgruber,
  {et~al.}}]{de1996sws}
De~Graauw, T., Whittet, D., Gerakines, P.~a., {et~al.} 1996, A\&A, 315, L345

\bibitem[{d'Hendecourt \& Jourdain~de Muizon(1989)}]{d1989discovery}
d'Hendecourt, L. \& Jourdain~de Muizon, M. 1989, A\&A, 223, L5

\bibitem[{Ehrenfreund {et~al.}(1997)Ehrenfreund, Boogert, Gerakines, Tielens,
  Van~Dishoeck, {et~al.}}]{ehrenfreund1997infrared}
Ehrenfreund, P., Boogert, A., Gerakines, P., {et~al.} 1997, A\&A, 328, 649

\bibitem[{Favre {et~al.}(2018)Favre, Fedele, Semenov, Parfenov, Codella,
  Ceccarelli, Bergin, Chapillon, Testi, Hersant, {et~al.}}]{favre2018first}
Favre, C., Fedele, D., Semenov, D., {et~al.} 2018, ApJL, 862, L2

\bibitem[{Fayolle {et~al.}(2011)Fayolle, {\"O}berg, Cuppen, Visser, \&
  Linnartz}]{fayolle2011laboratory}
Fayolle, E.~C., {\"O}berg, K., Cuppen, H.~M., Visser, R., \& Linnartz, H. 2011,
  A\&A, 529, A74

\bibitem[{Fedoseev {et~al.}(2017)Fedoseev, Chuang, Ioppolo, Qasim, van
  Dishoeck, \& Linnartz}]{fedoseev2017formation}
Fedoseev, G., Chuang, K.-J., Ioppolo, S., {et~al.} 2017, ApJ, 842, 52

\bibitem[{Francisco {et~al.}(2010)Francisco, Muckerman, \&
  Yu}]{francisco2010hoco}
Francisco, J.~S., Muckerman, J.~T., \& Yu, H.-G. 2010, Acc. Chem. Res, 43, 1519

\bibitem[{Fredon {et~al.}(2017)Fredon, Lamberts, \& Cuppen}]{fredon2017energy}
Fredon, A., Lamberts, T., \& Cuppen, H. 2017, ApJ, 849, 125

\bibitem[{Fuchs {et~al.}(2009)Fuchs, Cuppen, Ioppolo,
  {et~al.}}]{fuchs2009hydrogenation}
Fuchs, G., Cuppen, H., Ioppolo, S., {et~al.} 2009, A\&A, 505, 629

\bibitem[{Garrod \& Herbst(2006)}]{garrod2006formation}
Garrod, R.~T. \& Herbst, E. 2006, A\&A, 457, 927

\bibitem[{Gerakines {et~al.}(1996)Gerakines, Schutte, \&
  Ehrenfreund}]{gerakines1996ultraviolet}
Gerakines, P., Schutte, W., \& Ehrenfreund, P. 1996, A\&A, 312, 289

\bibitem[{Gerakines {et~al.}(1999)Gerakines, Whittet, Ehrenfreund, Boogert,
  Tielens, Schutte, Chiar, Van~Dishoeck, Prusti, Helmich,
  {et~al.}}]{gerakines1999observations}
Gerakines, P., Whittet, D., Ehrenfreund, P., {et~al.} 1999, ApJ, 522, 357

\bibitem[{Gibb {et~al.}(2004)Gibb, Whittet, Boogert, \&
  Tielens}]{gibb2004interstellar}
Gibb, E., Whittet, D., Boogert, A., \& Tielens, A. 2004, ApJS, 151, 35

\bibitem[{Giguere \& Harvey(1959)}]{giguere1959infrared}
Giguere, P. \& Harvey, K. 1959, J. Mol. Spectrosc., 3, 36

\bibitem[{Goumans \& Andersson(2010)}]{goumans2010tunnelling}
Goumans, T. \& Andersson, S. 2010, MNRAS, 406, 2213

\bibitem[{He {et~al.}(2017)He, Emtiaz, \& Vidali}]{he2017mechanism}
He, J., Emtiaz, S.~M., \& Vidali, G. 2017, ApJ, 851, 104

\bibitem[{Hidaka {et~al.}(2004)Hidaka, Watanabe, Shiraki, Nagaoka, \&
  Kouchi}]{hidaka2004conversion}
Hidaka, H., Watanabe, N., Shiraki, T., Nagaoka, A., \& Kouchi, A. 2004, ApJ,
  614, 1124

\bibitem[{Hodyss {et~al.}(2009)Hodyss, Johnson, Stern, Goguen, \&
  Kanik}]{hodyss2009photochemistry}
Hodyss, R., Johnson, P.~V., Stern, J.~V., Goguen, J.~D., \& Kanik, I. 2009,
  Icarus, 200, 338

\bibitem[{Hudson \& Moore(1999)}]{hudson1999laboratory}
Hudson, R. \& Moore, M. 1999, Icarus, 140, 451

\bibitem[{Ikeda {et~al.}(2001)Ikeda, Ohishi, Nummelin, Dickens, Bergman,
  Hjalmarson, \& Irvine}]{ikeda2001survey}
Ikeda, M., Ohishi, M., Nummelin, A., {et~al.} 2001, ApJ, 560, 792

\bibitem[{Ioppolo {et~al.}(2011{\natexlab{a}})Ioppolo, Cuppen, \&
  Linnartz}]{ioppolo2011surfacerend}
Ioppolo, S., Cuppen, H., \& Linnartz, H. 2011{\natexlab{a}}, Rend. Lincei, 22,
  211

\bibitem[{Ioppolo {et~al.}(2008)Ioppolo, Cuppen, Romanzin, Van~Dishoeck, \&
  Linnartz}]{ioppolo2008laboratory}
Ioppolo, S., Cuppen, H., Romanzin, C., Van~Dishoeck, E., \& Linnartz, H. 2008,
  ApJ, 686, 1474

\bibitem[{Ioppolo {et~al.}(2010)Ioppolo, Cuppen, Van~Dishoeck, \&
  Linnartz}]{ioppolo2010surface}
Ioppolo, S., Cuppen, H., Van~Dishoeck, E., \& Linnartz, H. 2010, MNRAS, 410,
  1089

\bibitem[{Ioppolo {et~al.}(2013{\natexlab{a}})Ioppolo, Fedoseev, Lamberts,
  Romanzin, \& Linnartz}]{ioppolo2013surfreside2}
Ioppolo, S., Fedoseev, G., Lamberts, T., Romanzin, C., \& Linnartz, H.
  2013{\natexlab{a}}, Rev. Sci. Instrum., 84, 073112

\bibitem[{Ioppolo {et~al.}(2013{\natexlab{b}})Ioppolo, Sangiorgio, Baratta, \&
  Palumbo}]{ioppolo2013solid}
Ioppolo, S., Sangiorgio, I., Baratta, G., \& Palumbo, M. 2013{\natexlab{b}},
  A\&A, 554, A34

\bibitem[{Ioppolo {et~al.}(2011{\natexlab{b}})Ioppolo, Van~Boheemen, Cuppen,
  Van~Dishoeck, \& Linnartz}]{ioppolo2011surface}
Ioppolo, S., Van~Boheemen, Y., Cuppen, H., Van~Dishoeck, E., \& Linnartz, H.
  2011{\natexlab{b}}, MNRAS, 413, 2281

\bibitem[{Irvine {et~al.}(1990)Irvine, Friberg, Kaifu, Matthews, Minh, Ohishi,
  \& Ishikawa}]{irvine1990detection}
Irvine, W.~M., Friberg, P., Kaifu, N., {et~al.} 1990, A\&A, 229, L9

\bibitem[{Jamieson {et~al.}(2006)Jamieson, Mebel, \&
  Kaiser}]{jamieson2006understanding}
Jamieson, C.~S., Mebel, A.~M., \& Kaiser, R.~I. 2006, ApJS, 163, 184

\bibitem[{Keane {et~al.}(2001)Keane, Tielens, Boogert, Schutte, \&
  Whittet}]{keane2001ice}
Keane, J., Tielens, A., Boogert, A., Schutte, W., \& Whittet, D. 2001, A\&A,
  376, 254

\bibitem[{Kim {et~al.}(2012)Kim, Evans~II, Dunham, Lee, \&
  Pontoppidan}]{kim2012co2}
Kim, H.~J., Evans~II, N.~J., Dunham, M.~M., Lee, J.-E., \& Pontoppidan, K.~M.
  2012, ApJ, 758, 38

\bibitem[{Lamberts {et~al.}(2013)Lamberts, Cuppen, Ioppolo, \&
  Linnartz}]{lamberts2013water}
Lamberts, T., Cuppen, H.~M., Ioppolo, S., \& Linnartz, H. 2013, Phys. Chem.
  Chem. Phys., 15, 8287

\bibitem[{Lamberts {et~al.}(2014)Lamberts, de~Vries, \&
  Cuppen}]{lamberts2014formation}
Lamberts, T., de~Vries, X., \& Cuppen, H. 2014, Faraday Discuss., 168, 327

\bibitem[{Lamberts {et~al.}(2017)Lamberts, Fedoseev, K{\"a}stner, Ioppolo, \&
  Linnartz}]{lamberts2017importance}
Lamberts, T., Fedoseev, G., K{\"a}stner, J., Ioppolo, S., \& Linnartz, H. 2017,
  A\&A, 599, A132

\bibitem[{Lamberts \& K{\"a}stner(2017)}]{lamberts2017influence}
Lamberts, T. \& K{\"a}stner, J. 2017, ApJ, 846, 43

\bibitem[{Lamberts {et~al.}(2016)Lamberts, Samanta, K{\"o}hn, \&
  K{\"a}stner}]{lamberts2016quantum}
Lamberts, T., Samanta, P.~K., K{\"o}hn, A., \& K{\"a}stner, J. 2016, Phys.
  Chem. Chem. Phys., 18, 33021

\bibitem[{Lannon {et~al.}(1971)Lannon, Verderame, \&
  Anderson~Jr}]{lannon1971infrared}
Lannon, J.~A., Verderame, F.~D., \& Anderson~Jr, R.~W. 1971, J. Chem. Phys.,
  54, 2212

\bibitem[{Linnartz {et~al.}(2015)Linnartz, Ioppolo, \&
  Fedoseev}]{linnartz2015atom}
Linnartz, H., Ioppolo, S., \& Fedoseev, G. 2015, Int. Rev. Phys. Chem., 34, 205

\bibitem[{Liu {et~al.}(2002)Liu, Girart, Remijan, \& Snyder}]{liu2002formic}
Liu, S.-Y., Girart, J., Remijan, A., \& Snyder, L. 2002, ApJ, 576, 255

\bibitem[{Liu {et~al.}(2001)Liu, Mehringer, \& Snyder}]{liu2001observations}
Liu, S.-Y., Mehringer, D.~M., \& Snyder, L.~E. 2001, ApJ, 552, 654

\bibitem[{Loeffler {et~al.}(2005)Loeffler, Baratta, Palumbo, Strazzulla, \&
  Baragiola}]{loeffler2005co}
Loeffler, M., Baratta, G., Palumbo, M., Strazzulla, G., \& Baragiola, R. 2005,
  A\&A, 435, 587

\bibitem[{Loeffler {et~al.}(2006)Loeffler, Teolis, \&
  Baragiola}]{loeffler2006decomposition}
Loeffler, M.~J., Teolis, B.~D., \& Baragiola, R.~A. 2006, J. Chem. Phys., 124,
  104702

\bibitem[{Maity {et~al.}(2014)Maity, Kaiser, \& Jones}]{maity2014infrared}
Maity, S., Kaiser, R.~I., \& Jones, B.~M. 2014, Faraday Discuss., 168, 485

\bibitem[{Martin {et~al.}(2008)Martin, Bertin, Domaracka, Azria, Illenberger,
  \& Lafosse}]{martin2008chemistry}
Martin, I., Bertin, M., Domaracka, A., {et~al.} 2008, Int. J. Mass Spectrom.,
  277, 262

\bibitem[{Mart{\'\i}n-Dom{\'e}nech {et~al.}(2015)Mart{\'\i}n-Dom{\'e}nech,
  Manzano-Santamar{\'\i}a, Caro, Cruz-D{\'\i}az, Chen, Herrero, \&
  Tanarro}]{martin2015uv}
Mart{\'\i}n-Dom{\'e}nech, R., Manzano-Santamar{\'\i}a, J., Caro, G.~M.,
  {et~al.} 2015, A\&A, 584, A14

\bibitem[{Masunov {et~al.}(2016)Masunov, Wait, \& Vasu}]{masunov2016chemical}
Masunov, A.~E., Wait, E., \& Vasu, S.~S. 2016, J. Phys. Chem. A, 120, 6023

\bibitem[{McCarthy {et~al.}(2016)McCarthy, Martinez~Jr, McGuire, Crabtree,
  Martin-Drumel, \& Stanton}]{mccarthy2016isotopic}
McCarthy, M.~C., Martinez~Jr, O., McGuire, B.~A., {et~al.} 2016, J. Chem.
  Phys., 144, 124304

\bibitem[{Meisner {et~al.}(2017)Meisner, Lamberts, \&
  K{\"a}stner}]{meisner2017atom}
Meisner, J., Lamberts, T., \& K{\"a}stner, J. 2017, ACS Earth and Space
  Chemistry, 1, 399

\bibitem[{Minissale {et~al.}(2015)Minissale, Loison, Baouche, Chaabouni,
  Congiu, \& Dulieu}]{minissale2015solid}
Minissale, M., Loison, J.-C., Baouche, S., {et~al.} 2015, A\&A, 577, A2

\bibitem[{Moore {et~al.}(1996)Moore, Ferrante, \& Nuth~III}]{moore1996infrared}
Moore, M., Ferrante, R., \& Nuth~III, J. 1996, Planet. Space Sci., 44, 927

\bibitem[{Mo{\.z}ejko(2007)}]{mozejko2007calculations}
Mo{\.z}ejko, P. 2007, Eur. Phys. J. Spec. Top., 144, 233

\bibitem[{Nguyen {et~al.}(2012)Nguyen, Xue, Weston~Jr, Barker, \&
  Stanton}]{nguyen2012reaction}
Nguyen, T.~L., Xue, B.~C., Weston~Jr, R.~E., Barker, J.~R., \& Stanton, J.~F.
  2012, J. Phys. Chem. Lett., 3, 1549

\bibitem[{Noble {et~al.}(2011)Noble, Dulieu, Congiu, \& Fraser}]{noble2011co2}
Noble, J.~A., Dulieu, F., Congiu, E., \& Fraser, H.~J. 2011, ApJ, 735, 121

\bibitem[{Oba {et~al.}(2010{\natexlab{a}})Oba, Watanabe, Kouchi, Hama, \&
  Pirronello}]{oba2010experimental}
Oba, Y., Watanabe, N., Kouchi, A., Hama, T., \& Pirronello, V.
  2010{\natexlab{a}}, ApJL, 712, L174

\bibitem[{Oba {et~al.}(2010{\natexlab{b}})Oba, Watanabe, Kouchi, Hama, \&
  Pirronello}]{oba2010formation}
Oba, Y., Watanabe, N., Kouchi, A., Hama, T., \& Pirronello, V.
  2010{\natexlab{b}}, ApJ, 722, 1598

\bibitem[{{\"O}berg {et~al.}(2011){\"O}berg, Boogert, Pontoppidan, Van~den
  Broek, Van~Dishoeck, Bottinelli, Blake, \& Evans~II}]{oberg2011spitzer}
{\"O}berg, K.~I., Boogert, A.~A., Pontoppidan, K.~M., {et~al.} 2011, ApJ, 740,
  109

\bibitem[{{\"O}berg {et~al.}(2009){\"O}berg, Garrod, Van~Dishoeck, \&
  Linnartz}]{oberg2009formation}
{\"O}berg, K.~I., Garrod, R.~T., Van~Dishoeck, E.~F., \& Linnartz, H. 2009,
  A\&A, 504, 891

\bibitem[{Orient \& Strivastava(1987)}]{orient1987electron}
Orient, O. \& Strivastava, S. 1987, J. Phys. B, 20, 3923

\bibitem[{Paardekooper {et~al.}(2016)Paardekooper, Bossa, \&
  Linnartz}]{paardekooper2016laser}
Paardekooper, D., Bossa, J.-B., \& Linnartz, H. 2016, A\&A, 592, A67

\bibitem[{Palumbo {et~al.}(1998)Palumbo, Baratta, Brucato, Castorina, Satorre,
  \& Strazzulla}]{palumbo1998profile}
Palumbo, M., Baratta, G., Brucato, J., {et~al.} 1998, A\&A, 334, 247

\bibitem[{Pontoppidan(2006)}]{pontoppidan2006spatial}
Pontoppidan, K.~M. 2006, A\&A, 453, L47

\bibitem[{Pontoppidan {et~al.}(2008)Pontoppidan, Boogert, Fraser, van Dishoeck,
  Blake, Lahuis, {\"O}berg, Evans~II, \& Salyk}]{pontoppidan2008c2d}
Pontoppidan, K.~M., Boogert, A.~C., Fraser, H.~J., {et~al.} 2008, ApJ, 678,
  1005

\bibitem[{Poteet {et~al.}(2013)Poteet, Pontoppidan, Megeath, Watson, Isokoski,
  Bjorkman, Sheehan, \& Linnartz}]{poteet2013anomalous}
Poteet, C.~A., Pontoppidan, K.~M., Megeath, S.~T., {et~al.} 2013, ApJ, 766, 117

\bibitem[{Qasim {et~al.}(2018)Qasim, Chuang, Fedoseev, Ioppolo, Boogert, \&
  Linnartz}]{qasim2018formation}
Qasim, D., Chuang, K.-J., Fedoseev, G., {et~al.} 2018, A\&A, 612, A83

\bibitem[{Raut \& Baragiola(2011)}]{raut2011solid}
Raut, U. \& Baragiola, R. 2011, ApJL, 737, L14

\bibitem[{Raut {et~al.}(2012)Raut, Fulvio, Loeffler, \&
  Baragiola}]{raut2012radiation}
Raut, U., Fulvio, D., Loeffler, M., \& Baragiola, R. 2012, ApJ, 752, 159

\bibitem[{Requena-Torres {et~al.}(2006)Requena-Torres, Mart{\'\i}n-Pintado,
  Rodr{\'\i}guez-Franco, Mart{\'\i}n, Rodr{\'\i}guez-Fern{\'a}ndez, \&
  De~Vicente}]{requena2006organic}
Requena-Torres, M.~A., Mart{\'\i}n-Pintado, J., Rodr{\'\i}guez-Franco, A.,
  {et~al.} 2006, A\&A, 455, 971

\bibitem[{Romanzin {et~al.}(2011)Romanzin, Ioppolo, Cuppen,
  {et~al.}}]{romanzin2011water}
Romanzin, C., Ioppolo, S., Cuppen, H., {et~al.} 2011, J. Chem. Phys., 134,
  084504

\bibitem[{Roser {et~al.}(2001)Roser, Vidali, Manic{\`o}, \&
  Pirronello}]{roser2001formation}
Roser, J.~E., Vidali, G., Manic{\`o}, G., \& Pirronello, V. 2001, ApJL, 555,
  L61

\bibitem[{Ryazantsev \& Feldman(2015)}]{ryazantsev2015radiation}
Ryazantsev, S.~V. \& Feldman, V.~I. 2015, Phys. Chem. Chem. Phys., 17, 30648

\bibitem[{Satorre {et~al.}(2000)Satorre, Palumbo, \&
  Strazzulla}]{satorre2000co}
Satorre, M., Palumbo, M., \& Strazzulla, G. 2000, Astrophys. Space Sci., 274,
  643

\bibitem[{Schutte {et~al.}(1999)Schutte, Boogert, Tielens, Whittet, Gerakines,
  Chiar, Ehrenfreund, Greenberg, Van~Dishoeck, Graauw,
  {et~al.}}]{schutte1999weak}
Schutte, W., Boogert, A., Tielens, A., {et~al.} 1999, A\&A, 343, 966

\bibitem[{Song \& K{\"a}stner(2017)}]{song2017tunneling}
Song, L. \& K{\"a}stner, J. 2017, ApJ, 850, 118

\bibitem[{Strazzulla {et~al.}(1997)Strazzulla, Brucato, Palumbo, \&
  Satorre}]{strazzulla1997possible}
Strazzulla, G., Brucato, J., Palumbo, M., \& Satorre, M. 1997, A\&A, 321, 618

\bibitem[{Suhasaria {et~al.}(2017)Suhasaria, Baratta, Ioppolo, Zacharias, \&
  Palumbo}]{suhasaria2017solid}
Suhasaria, T., Baratta, G., Ioppolo, S., Zacharias, H., \& Palumbo, M. 2017,
  A\&A, 608, A12

\bibitem[{Tachikawa \& Kawabata(2016)}]{tachikawa2016effects}
Tachikawa, H. \& Kawabata, H. 2016, J. Phys. Chem. A, 120, 6596

\bibitem[{Taquet {et~al.}(2017)Taquet, Wirstr{\"o}m, Charnley, Faure,
  L{\'o}pez-Sepulcre, \& Persson}]{taquet2017chemical}
Taquet, V., Wirstr{\"o}m, E., Charnley, S.~B., {et~al.} 2017, A\&A, 607, A20

\bibitem[{Tielens {et~al.}(1982)Tielens, Hagen, {et~al.}}]{tielens1982model}
Tielens, A., Hagen, W., {et~al.} 1982, A\&A, 114, 245

\bibitem[{Trottier \& Brooks(2004)}]{trottier2004carbon}
Trottier, A. \& Brooks, R.~L. 2004, ApJ, 612, 1214

\bibitem[{Tschersich(2000)}]{tschersich2000intensity}
Tschersich, K. 2000, J. Appl. Phys., 87, 2565

\bibitem[{Tschersich {et~al.}(2008)Tschersich, Fleischhauer, \&
  Schuler}]{tschersich2008design}
Tschersich, K., Fleischhauer, J., \& Schuler, H. 2008, J. Appl. Phys., 104,
  034908

\bibitem[{Tschersich \& Von~Bonin(1998)}]{tschersich1998formation}
Tschersich, K. \& Von~Bonin, V. 1998, J. Appl. Phys., 84, 4065

\bibitem[{Turner {et~al.}(1999)Turner, Terzieva, \& Herbst}]{turner1999physics}
Turner, B., Terzieva, R., \& Herbst, E. 1999, ApJ, 518, 699

\bibitem[{Van~Dishoeck {et~al.}(1996)Van~Dishoeck, Helmich, de~Graauw, Black,
  Boogert, Ehrenfreund, Gerakines, Lacy, Millar, Schutte,
  {et~al.}}]{van1996search}
Van~Dishoeck, E., Helmich, F., de~Graauw, T., {et~al.} 1996, A\&A, 315, L349

\bibitem[{van Dishoeck(1998)}]{van1998can}
van Dishoeck, E.~F. 1998, Faraday Discuss., 109, 31

\bibitem[{Vandenbussche {et~al.}(1999)Vandenbussche, Ehrenfreund, Boogert,
  Van~Dishoeck, Schutte, Gerakines, Chiar, Tielens, Keane, Whittet,
  {et~al.}}]{vandenbussche1999constraints}
Vandenbussche, B., Ehrenfreund, P., Boogert, A., {et~al.} 1999, A\&A, 346, L57

\bibitem[{Watanabe \& Kouchi(2002)}]{watanabe2002efficient}
Watanabe, N. \& Kouchi, A. 2002, ApJL, 571, L173

\bibitem[{Watanabe {et~al.}(2007)Watanabe, Mouri, Nagaoka, Chigai, Kouchi, \&
  Pirronello}]{watanabe2007laboratory}
Watanabe, N., Mouri, O., Nagaoka, A., {et~al.} 2007, ApJ, 668, 1001

\bibitem[{Watanabe {et~al.}(2003)Watanabe, Shiraki, \&
  Kouchi}]{watanabe2003dependence}
Watanabe, N., Shiraki, T., \& Kouchi, A. 2003, ApJL, 588, L121

\bibitem[{Whittet {et~al.}(1998)Whittet, Gerakines, Tielens, Adamson, Boogert,
  Chiar, de~Graauw, Ehrenfreund, Prusti, Schutte,
  {et~al.}}]{whittet1998detection}
Whittet, D., Gerakines, P., Tielens, A., {et~al.} 1998, ApJL, 498, L159

\bibitem[{Winnewisser \& Churchwell(1975)}]{winnewisser1975detection}
Winnewisser, G. \& Churchwell, E. 1975, ApJ, 200, L33

\bibitem[{Yu {et~al.}(2005)Yu, Muckerman, \& Francisco}]{yu2005direct}
Yu, H.-G., Muckerman, J.~T., \& Francisco, J.~S. 2005, J. Phys. Chem. A, 109,
  5230

\bibitem[{Zanchet {et~al.}(2018)Zanchet, del Mazo, Aguado, Roncero,
  Jim{\'e}nez, Canosa, Ag{\'u}ndez, \& Cernicharo}]{zanchet2018full}
Zanchet, A., del Mazo, P., Aguado, A., {et~al.} 2018, Phys. Chem. Chem. Phys.,
  20, 5415

\bibitem[{Zuckerman {et~al.}(1971)Zuckerman, Ball, \&
  Gottlieb}]{zuckerman1971microwave}
Zuckerman, B., Ball, J.~A., \& Gottlieb, C.~A. 1971, ApJ, 163, L41

\end{thebibliography}

\end{document}